\documentclass[authoryear,preprint,12pt]{elsarticle}

\usepackage{amssymb}

\usepackage{amsmath}
\usepackage{float}
\usepackage{caption}
\usepackage{subcaption}
\usepackage{xcolor}
\usepackage{tabularx}
\usepackage{hyperref}
\usepackage{adjustbox}

\usepackage{algorithm}
\usepackage{algpseudocode}

\journal{Computer Methods in Applied Mechanics and Engineering}

\begin{document}

\begin{frontmatter}



\title{An Adaptive Framework for Autoregressive Forecasting in CFD Using Hybrid Modal Decomposition and Deep Learning}


\author[inst1]{Rodrigo Abadía-Heredia}
\author[inst1]{Manuel Lopez-Martin}
\author[inst1]{Soledad Le Clainche}

\affiliation[inst1]{
    organization={ETSI Aeronáutica y del Espacio, Universidad Politécnica de Madrid},
    addressline={Plaza Cardenal Cisneros, 3}, 
    city={Madrid},
    postcode={28040}, 
    state={Madrid},
    country={Spain}
}

\begin{abstract}
This work presents, to the best of the authors’ knowledge, the first generalizable and fully data-driven adaptive framework designed to stabilize deep learning (DL) autoregressive forecasting models over long time horizons, with the goal of reducing the computational cost required in computational fluid dynamics (CFD) simulations.The proposed methodology alternates between two phases: (i) predicting the evolution of the flow field over a selected time interval using a trained DL model, and (ii) updating the model with newly generated CFD data when stability degrades, thus maintaining accurate long-term forecasting. This adaptive retraining strategy ensures robustness while avoiding the accumulation of predictive errors typical in autoregressive models. The framework is validated across three increasingly complex flow regimes, from laminar to turbulent, demonstrating from $30\%$ to $95\%$ reduction in computational cost without compromising physical consistency or accuracy. Its entirely data-driven nature makes it easily adaptable to a wide range of time-dependent simulation problems. The code implementing this methodology is available as open-source\footnote{https://github.com/RAbadiaH/adaptive-cfd-forecasting-hybrid-modal-decomposition-deep-learning} and it will be integrated into the upcoming release of the ModelFLOWs-app\footnote{https://modelflows.github.io/modelflowsapp/}.
\end{abstract}




\end{frontmatter}


\section{Introduction}\label{sec: introduction}
In fluid dynamics numerical simulations are fundamental in both industrial and academic contexts for predicting and analyzing complex fluid flow phenomena. High-fidelity simulations, such as Direct Numerical Simulation (DNS) or Large Eddy Simulations (LES), are often required to capture the full range of spatio-temporal scales in the flow dynamics evolving in time. However, these strategies are computationally intensive and time-consuming, making them impractical for many real-world applications. This challenge underscores the need for methodologies that can significantly reduce computational costs while preserving, to a reasonable extent, the accuracy and reliability associated with DNS or LES.

In this context, reduced-order models (ROMs) have emerged as a viable alternative to numerical simulations, providing a way to significantly lower computational costs while maintaining an acceptable level of accuracy. ROMs are typically classified into two main categories: intrusive and non-intrusive models.

Intrusive ROMs involve direct manipulation of the governing equations of the physical system. A prominent example of this approach are the Galerkin methods, where the governing equations are projected onto a reduced-order basis. This projection yields a simplified system of equations that significantly reduces the computational effort required for their solution \cite{Rempfer2000}.

In contrast, non-intrusive ROMs, often referred to as surrogate or data-driven models, do not require any modification of the governing equations. Instead, they rely exclusively on data to infer and predict the system’s dynamics. Within this framework, machine learning models are widely adopted for their ability to capture complex and nonlinear relationships directly from data, enabling accurate predictions without explicit knowledge of the underlying physical laws \cite{2018_vlachas_etal_forecasting_chaos, 2019_han_etal_forecasting_convAutoEnc, 2020_hasegawa_etal_forecasting_convAutoEnc_2D, 2021_nakamura_etal_forecasting_convAutoEnc_3D, 2024_gao_etal_gled}.

Galerkin methods have been extensively used to simulate flow dynamics by projecting the governing equations onto an orthonormal basis constructed from proper orthogonal decomposition (POD) modes \cite{rapun_adaptive_pod_modes_2015}. This approach shifts the focus from solving the full system to simulating the temporal evolution of mode amplitudes, thereby reducing the dimensionality and complexity of the original problem \cite{2017_leClainche_etal_adaptive_oil}. Despite its effectiveness in various contexts, Galerkin methods faces notable challenges. One key issue is that they can produce unstable iteration schemes \cite{galerkin_problem}.

Furthermore, in some cases the flow dynamics cannot be accurately captured using only a few dominant POD modes. This limitation necessitates the inclusion of a larger number of modes to preserve the essential structures of the flow \cite{galerkin_low_num_modes}, which in turn increases the computational cost and may compromise the efficiency gains of the ROM.

In this context, data-driven models have emerged as a compelling alternative for predicting the evolution of mode amplitudes without requiring projection of the governing equations. By combining such models with POD, a hybrid methodology is established that leverages the strengths of both dimensionality reduction and data-based learning. This approach is usually known as POD-DL and has been extensively studied for its effectiveness in modeling the temporal evolution of dynamical systems \cite{forecasting_pod_dl_2018, forecasting_pod_dl_2019, pod_forecating_2019_regazzoni, pod_forecating_2020_parish, abadiaherediaetal_2022, forecasting_pod_dl_2022, corrochano_etal2023_NNComb, 2025_abadiaheredia_etal_poddl_ae, pod_dl_Besabe_etal_2025}.

Deep learning (DL) models are frequently selected for this task due to their capability to model nonlinear dynamics, their relatively low computational cost during training, and their demonstrated success in time-series forecasting \cite{2020_parish_etal_LSTM_traditional}. These models are adept at capturing intricate temporal dependencies and interactions in data, making them particularly effective for predicting the evolution of fluid flows \cite{2025_sengupta_eta_hybridML}.

However, a significant challenge associated with data-driven models is in managing the divergence of predictions from the reference dynamics, specially when they are computed in an autoregressive manner. This divergence often prevents models from accurately covering the full temporal horizon over which the system is intended to be solved \cite{adaled_petros, 2025_abadiaheredia_etal_poddl_ae}.

The proposed solutions generally fall into two categories: one focuses on the development of predictive models that exhibit greater robustness to divergence, such as diffusion models \cite{Lin_etal_2024_diffusion_forecasting_survey}, while the other centers on adaptive frameworks that integrate classical numerical simulation techniques with data-driven approaches to dynamically correct and guide the predictions \cite{beltran_adaptive, 2021_leClainche_hodmd_book}.

This work focuses on the latter approach, wherein adaptive frameworks combine numerical solvers with deep learning models to enhance predictive performance. There is considerable interest in combining DL models with numerical solvers, particularly for high-dimensional and complex problems, where training a DL model, and computing the corresponding predictions can be significantly faster than the computation time required by a numerical solver for a single time step \cite{adaled_petros}. Additionally, numerical solvers often require multiple intermediate time steps, which are typically not stored, to compute the subsequent time step of interest. In contrast, DL models can directly predict the next time step of interest without the need for intermediate computations, thereby reducing computational overhead.

Several studies have proposed such adaptive strategies. For instance, \cite{adaled_petros} introduced a framework that integrates a numerical solver with a purely DL model to approximate solutions of various dynamical systems. In this method, the numerical solver is first used to generate high-fidelity solutions, which serve as training data for the DL model. Once trained, the DL model forecasts future states until its predictions begin to diverge. At that point, the numerical solver is re-invoked to compute new accurate solutions, which are subsequently used to retrain the DL model. This cycle is repeated, allowing the framework to adapt dynamically over the temporal domain.

A similar approach that integrates physics-based simulators with machine learning models was proposed by \cite{2024_weather_forecast} in the context of weather forecasting, where it achieved state-of-the-art performance for forecasts extending up to fifteen days.

Another strategy replaces the use of DL models with higher-order dynamic mode decomposition (HODMD) \cite{2017_leClainche_etal_hodmd} to approximate the solution of the complex Ginzburg–Landau equation \cite{beltran_adaptive} and also to predict complex problems in fluid dynamics \cite{2019_beltran_etal_soco}.

This work presents, for the first time to the authors knowledge, an adaptive framework similar to those previously introduced, in which the predictive model is the hybrid POD-DL model. The novelty of this approach lies in leveraging the strengths of this hybrid architecture, which has been shown to improve both the accuracy and stability of predictions when compared to purely DL-based models, while also substantially reducing training time, as demonstrated in \cite{2025_abadiaheredia_etal_poddl_ae}. This efficiency gain is particularly critical for adaptive frameworks designed to reduce the computational cost relative to conventional numerical solvers.

Although the POD-DL model exhibits greater stability compared to purely DL-based models, its autoregressive nature leads to the accumulation of errors over time, as inaccuracies in earlier predictions are propagated and amplified in subsequent forecasts. Making necessary the implementation of an adaptive framework that updates the POD-DL model with new information.

This work is organized as follows. Section \ref{sec: adaptive_framework} provides a detailed description of the adaptive framework proposed in this study. Section \ref{sec: dl_model} describes the POD-DL model employed to predict the flow dynamics. The results obtained by applying the adaptive framework to precomputed datasets are presented in Section \ref{sec: results}, and a general discussion of these findings is given in Section \ref{sec: discussion}. Finally, Section \ref{sec: conclusions} summarizes the main conclusions drawn from this work.

\section{Adaptive framework} \label{sec: adaptive_framework}

The dynamics of fluid flow are typically governed by a system of nonlinear partial differential equations (PDEs). In most cases, obtaining solutions to these equations necessitates numerical methods, except for a few special cases where analytical solutions are readily available.

The resolution of this system of equations can be generally described as follows. Let's suppose that $\mathcal{P}_{t} \subset \mathbb{R}$ and $\mathcal{P}_{x} \subset \mathbb{R}^{3}$ (or $\mathbb{R}^{2}$ for two-dimensional cases) are compact sets representing the temporal and spatial domains, respectively, where the system of PDEs is to be solved.

Let $\mathcal{P} = (\mathcal{P}_{x}, \mathcal{P}_{t})$ denote the combined spatio-temporal domain. Consider a Hilbert space $\mathcal{H} = \mathcal{H}(\mathcal{P})$ defined over $\mathcal{P}$, which represents the space of all possible solutions to the system of PDEs within this domain. Specifically, $\mathcal{H}$ consists of functions $\boldsymbol{v}$ such that $\boldsymbol{v}: \mathcal{P}_{x} \times \mathcal{P}_{t} \rightarrow \mathbb{R}$. Furthermore, the system of nonlinear PDEs can be expressed as a mapping $\Phi: \mathcal{H} \times \mathcal{P} \rightarrow \mathbb{R}$. The problem of interest can then be stated as follows: given a point $p = (x,t) \in \mathcal{P} = (\mathcal{P}_{x}, \mathcal{P}_{t})$, find a solution $\boldsymbol{v}(p) \in \mathcal{H}(\mathcal{P})$ of the governing equations represented by $\Phi$ such that,

\begin{equation}
    \Phi(\boldsymbol{v}(p), p) = 0.
    \label{eq: pde}
\end{equation}

In practice, the goal is to find a function $\boldsymbol{v} \in \mathcal{H}$ that minimizes the deviation of eq. \eqref{eq: pde} from zero, rather than achieving an exact zero. Furthermore, to approximate the solution, it is necessary to discretize the governing eq. \eqref{eq: pde} in both time and space.

This significantly increases the degrees of freedom to the order of thousands or even millions, thereby escalating the computational cost to a level where solving the equations often requires specialized hardware and/or extended computation times.

In certain scenarios, resolving all spatio-temporal scales of the flow dynamics is computationally prohibitive. In such cases, reduced-order models (ROMs) provide a practical solution by approximating the dynamics for only the scales of interest, since in many practical scenarios only the large scales are required \cite{Jimenez_2018}. For instance, in Galerkin methods, the governing equations are projected onto an orthonormal basis constructed from the most energetic proper orthogonal decomposition (POD) modes \cite{sirovich_svd}.

However, this reduction in complexity by focusing on the flow scales of interest, may still require significant computational time. This study explores an adaptive methodology that leverages data obtained from classical numerical solvers in combination with deep learning (DL) forecasting models to alleviate the computational cost associated with computing these solutions at all required time steps.

More specifically this adaptive methodology is used to approximate the solution, $\boldsymbol{v}$, over the domain $\mathcal{P} = (\mathcal{P}_{x}, \mathcal{P}_{t})$, where $\mathcal{P}_{t} \subset [0, T]$. In an ideal implementation of the adaptive methodology, a live numerical solver is employed to approximate the solution. However, note that the methodology developed is fully data-driven, where the CFD solver can be replaced by a group of snapshots, that can be calculated numerically or experimentally. The adaptive framework is as follows,
\begin{enumerate}
    \item The numerical solver computes a set of solutions, $\boldsymbol{\mathring{D}} = \{\boldsymbol{v}_{0}, \dots, \boldsymbol{v}_{T_{1}}\}$, for time steps $\{0, \dots, T_{1}\}$, where $T_{1} \ll T$.
    \item This dataset $\boldsymbol{\mathring{D}}$ is used to train a DL forecasting model $\mathcal{M}$, which predicts the solutions for the subsequent time steps, i.e., $\mathcal{M}(\boldsymbol{\mathring{D}}) = \{\boldsymbol{v}_{T_{1} + 1}, \dots, \boldsymbol{v}_{T_{1} + T_{2}}\}$.
    \item The numerical solver is then invoked again to compute additional solutions, $\boldsymbol{\mathring{D}} = \{\boldsymbol{v}_{T_{1} + T_{2} + 1}, \dots, \boldsymbol{v}_{T_{1} + T_{2} + T_{3}}\}$.
    \item Repeat steps (2) and (3) until the solution spans the entire temporal domain up to $T$.
\end{enumerate}

Since this work is intended as an initial exploration of the adaptive framework, the focus is placed on assessing its effectiveness in addressing complex problems in fluid dynamics. In this context, this study aims to evaluate both the theoretical speed-up that the framework could offer and the potential challenges encountered by the underlying DL forecasting model.

Therefore, instead of employing a live numerical solver, it is substituted with a precomputed dataset. Each time the methodology calls for new solutions from the numerical solver (step 3), they are retrieved from the precomputed dataset. This modification allows to focus on evaluating the performance of the DL forecasting model in this adaptive framework without the additional computational overhead of real-time numerical computations. The integration of a live numerical solver is left open for future research.

The following section provides a detailed explanation of the DL forecasting model utilized in this work, including its motivation, architecture and the training procedure employed.

\section{Forecasting model: POD-DL}\label{sec: dl_model}
The forecasting model used in this work to predict the future evolution of the dynamics is based on a hybrid model combining POD modes and DL architectures. As previously mentioned, this model has been already studied in the literature for a large variety problems \cite{forecasting_pod_dl_2018, forecasting_pod_dl_2019, pod_dl_interpolation, pod_forecating_2019_regazzoni, pod_forecating_2020_parish, abadiaherediaetal_2022, forecasting_pod_dl_2022, corrochano_etal2023_NNComb, 2025_abadiaheredia_etal_poddl_ae, pod_dl_Besabe_etal_2025} and is usually identified as POD-DL. This name will be used for the rest of this work.

\subsection{Proper orthogonal decomposition}\label{sec: pod}
To introduce this model, we begin with the POD, a technique that enables the projection of the solution of the governing equations \eqref{eq: pde} onto an orthonormal basis, expressed as,

\begin{equation}
    \boldsymbol{v}(x,t) = \sum_{i=1}^{K} \boldsymbol{a}_{i}(t) \boldsymbol{U}_{i}(x).
    \label{eq: pod}
\end{equation}

Here the set $\{\boldsymbol{U}_{1}, \dots, \boldsymbol{U}_{K}\}$ represents the orthonormal basis, or POD modes, and $\boldsymbol{a}_{i}$ denotes the mode amplitudes, which, when applied to velocity vectors, quantify the energy contribution of each POD mode to the overall dynamics \cite{aubry_1991}. These amplitudes facilitate the classification of POD modes based on the amount of information they capture about the dynamics. This technique has been widely used to analyze the structural composition of flow dynamics, enabling the identification of coherent structures within the flow \cite{holmes_lumley_libro, Mendez_2020_experimental, pod_40_energy_causality}. The application of POD to fluid dynamics was first introduced by \cite{lumley_pod}.

In this work, both the POD modes and their corresponding mode amplitudes are computed using singular value decomposition (SVD), commonly referred to as the method of snapshots, introduced by Sirovich in $1987$ \cite{sirovich_svd}.

This technique requires a dataset $\boldsymbol{\mathring{D}}$ conformed by a sequence of snapshots representing the flow dynamics, e.g., the velocity flow field, pressure field, etc. The resolution of these snapshots is related to the spatial grid used to discretize the spatial domain $\mathcal{P}_{x}$.

Given that the spatio-temporal domain $\mathcal{P} = (\mathcal{P}_{x}, \mathcal{P}_{t})$ is compact, we assume that the spatial domain is discretized into $N = \# \mathcal{P}_{x}$ points, $x_{j} \in \mathbb{R}^{3}$ (or $\mathbb{R}^{2}$). Let $S$ denote the number of available snapshots. The dataset $\boldsymbol{\mathring{D}}$ is then represented as a snapshot tensor, defined as follows,


\begin{equation} \label{eq: dataset}
    \boldsymbol{\mathring{D}} = [\boldsymbol{v}_{0}, \dots, \boldsymbol{v}_{S - 1}], \quad \boldsymbol{\mathring{D}} \in \mathbb{R}^{C \times N_{x} \times N_{y} \times N_{z} \times S}.
\end{equation}

Here $\boldsymbol{v}_{i} \equiv \boldsymbol{v}(\cdot, i)$ and $C$ represents the number of components, such as velocity components (streamwise, wall-normal and spanwise), different Reynolds numbers, or other relevant parameters. The variables $N_{x}$, $N_{y}$ and $N_{z}$ indicate the number of discretization points along the $x$-, $y$- and $z$-axis, respectively.

Note, this dataset is composed of solutions to the system of PDEs in eq. \eqref{eq: pde} within a subdomain $[0, T_{1}] \subset \mathcal{P}_{t} = [0, T]$, which represents a subset of the temporal domain of interest. Consequently, to generate the snapshots required for the application of SVD, it is necessary to solve the system in eq. \eqref{eq: pde} over a temporal horizon $T_{1}$ that is as small as feasible, ideally ensuring that $T_{1} \ll T$. 

Before applying SVD to the tensor $\boldsymbol{\mathring{D}}$, it has to be reshaped into a matrix $\boldsymbol{D}$, such that $\boldsymbol{D} \in \mathbb{R}^{C * N_{x} * N_{y} * N_{z} \times S}$. Ideally SVD is applied to the fluctuations in the dynamics, by subtracting the mean flow from the original dynamics, as follows,

\begin{align}
     \boldsymbol{\tilde{D}} & =  \boldsymbol{D} -  \boldsymbol{\bar{D}}, \\
    \boldsymbol{\bar{D}} & = \frac{1}{S} \sum_{i = 1}^{S} \boldsymbol{D}_{:,i}, \\
    \hbox{\textit{dim}}(\boldsymbol{\bar{D}}) & = [C * N_{x} * N_{y} * N_{z} \times 1].
\end{align}

Here $\boldsymbol{\tilde{D}}$ denotes the fluctuations within the dynamics, while $\boldsymbol{\bar{D}}$ represents the mean flow, capturing the steady structures of the system. Applying SVD to the fluctuation field results in the following matrix decomposition,

\begin{equation}
    {\boldsymbol{\tilde{D}}} = \boldsymbol{U} \boldsymbol{\Sigma} \boldsymbol{\Gamma^{*}},
    \label{eq: svd}
\end{equation}

where $\boldsymbol{U} \in \mathbb{R}^{C * N_{x} * N_{y} * N_{z} \times K}$ is a unitary matrix whose columns represent the POD modes in eq. \eqref{eq: pod}, $\boldsymbol{\Sigma} \in \mathbb{R}^{K \times K}$ is a diagonal matrix containing the singular values, and $\boldsymbol{\Gamma} \in \mathbb{R}^{S \times K}$ is a unitary matrix whose columns represent the POD coefficients. The operator $(\cdot)^{*}$ is the conjugate transpose operator. The tensor $\boldsymbol{\mathring{D}}$ can be recovered by adding the mean flow and inverting the reshape operation. Since $\boldsymbol{\Sigma}$ is diagonal, eq. \eqref{eq: svd} can be reformulated as a sum of rank-one matrices, expressed as,

\begin{equation}
    {\boldsymbol{\tilde{D}}} = \sum_{i = 1}^{K} \sigma_{i}\boldsymbol{u}_{i}\boldsymbol{\gamma}_{i}^{*}.
    \label{eq: svd_one_rank}
\end{equation}

Here $\sigma_{i} \in \mathbb{R}^{+}$, is non-negative and denotes the $i$-th singular value (i.e., the $i$-th element in the diagonal), $\boldsymbol{u}_{i} \equiv \boldsymbol{U}_{:,i} \in \mathbb{R}^{C * N_{x} * N_{y} * N_{z} \times 1}$ represents the $i$-th column in $\boldsymbol{U}$, which corresponds to the $i$-th POD mode, and $\boldsymbol{\gamma}_{i} \equiv \boldsymbol{\Gamma}_{:,i}$ is the $i$-th column of $\boldsymbol{\Gamma}$, representing the $i$-th POD coefficient. The variable $K$ defines the number of POD modes retained after truncation.

A key property of SVD is that the singular values are hierarchically ordered based on their energy contribution, such that $\sigma_{i} \geq \sigma_{j}$ for $i > j$. This implies that the $i$-th POD mode captures information about larger-scale structures within the dynamics compared to the $j$-th POD mode. This hierarchical structure enables the truncation of POD modes, retaining only those corresponding to the flow scales of interest while discarding the less significant ones \cite{pod_40_energy_causality}.

Note from eq. \eqref{eq: svd_one_rank} that the mode amplitudes in eq. \eqref{eq: pod} can be expressed as the product of the singular values and the POD coefficients,

\begin{equation}
    \boldsymbol{a}_{i}(t) = \sigma_{i}\boldsymbol{\gamma}_{t,i}.
    \label{eq: svd_amplitudes}
\end{equation}

Here $\gamma_{t,i} \equiv \boldsymbol{\Gamma}_{t,i}$ represents the $t$-th element of the $i$-th POD coefficient $\boldsymbol{\gamma}_{i}$. Once POD modes are computed using SVD, the temporal evolution of their corresponding amplitudes, $\boldsymbol{a}(t)$, can be predicted using data-driven models. As indicated in eq. \eqref{eq: pod}, predicting the future states of these amplitudes effectively forecasts the temporal evolution of the POD modes, which capture the key structures within the dynamics.

This methodology takes inspiration from Galerkin methods \cite{2009_luchtenburg_etal_galerkin, 2011_noack_etal_galerkin,2016_quarteroni_etal_galerkin}, where the governing eq. \eqref{eq: pde} are projected onto the orthonormal basis formed by the POD modes. These projected equations are expressed in terms of the POD amplitudes and solved using numerical methods \cite{rapun_adaptive_pod_modes_2015}. Therefore, Galerkin methods numerically approximate the solution of governing equations that describes the dynamics of POD modes. The flow dynamics are then recovered using eq. \eqref{eq: pod}.

In this work, the evolution of the POD modes is not derived by projecting the governing equations but is instead determined by applying SVD to an available dataset and using a data-driven model to forecast the POD coefficients $\boldsymbol{\gamma}_{i}$.

This approach enables the development of a non-intrusive ROM, as it does not require modifications to the governing equations. However, relying solely on a data-driven model limits its ability to generalize to unseen dynamics, such as changes in attractors or flow regimes. This limitation underscores the motivation to explore the integration of these fast data-driven models with numerical solvers within an adaptive framework.

Note that, since POD coefficients are represented as vectors, the complexity of the models required for forecasting is significantly reduced. This simplification makes it feasible to employ both traditional probabilistic forecasting methods \cite{var_models_tsay_2001} and DL models. In this work, a DL model was chosen over probabilistic approaches, guided by evidence from previous studies indicating that DL models consistently outperform traditional methods in forecasting tasks \cite{MAKRIDAKIS_etal_2018_M4, pod_forecating_2020_parish}. The next section describes the architecture of the DL model used in this work.

\subsection{Forecasting deep learning model}

The forecasting DL model, whose architecture is detailed in Tab. \ref{tab: pod_dl}, is designed to capture temporal correlations within a window of length $m$, constructed from a sequence of past POD coefficients, $\{ \boldsymbol{\gamma}^{s-m+1}, \dots, \boldsymbol{\gamma}^{s} \}$, and leverage this information to predict the future state of the coefficients, $\{ \boldsymbol{\gamma}^{s+1} \}$. Here, $\boldsymbol{\gamma}^{s} \equiv \boldsymbol{\Gamma}_{s,1:k}$ represents the $s$-th state (i.e., $s$-th snapshot) of the first $k$ retained coefficients. 

Note from eq. \eqref{eq: pod} that $k \leq K$, where $K$ denotes the total number of POD modes. If $k$ is substituted into eq. \eqref{eq: pod}, the resulting expression represents an approximation rather than an exact equality, as it excludes the contribution of the modes beyond the first $k$, thereby discarding part of the information contained in the original dynamics.

To perform the prediction, a long-short term memory (LSTM) \cite{hochreiter_etal_1997_lstm} architecture is used as the first layer of the model. This architecture is distinguished by its use of three gates: the input gate, forget gate, and output gate. They regulate the flow of information from the input sequences through the network. Specifically, the input gate selectively adds new information, the forget gate removes irrelevant information, and the output gate controls the transfer of information to the next cell in the sequence.

In this work, the input gate is denoted by $\boldsymbol{i}$, the forget gate by $\boldsymbol{f}$, and the output gate by $\boldsymbol{o}$. Additionally, the cell state is represented as $\boldsymbol{c}$, and the hidden state as $\boldsymbol{h}$. The computation of these gates and states follows the standard LSTM procedure, as described below,

\begin{align}
    \boldsymbol{f_{j}} & = \hat{\sigma}(\boldsymbol{W}_{f} [\boldsymbol{h}_{j-1}, \boldsymbol{x}_{j}] + \boldsymbol{b}_{f}) \\
    \boldsymbol{i}_{j} & = \hat{\sigma}(\boldsymbol{W}_{i} [\boldsymbol{h}_{j-1}, \boldsymbol{x}_{j}] + \boldsymbol{b}_{i}) \\
    \boldsymbol{\tilde{c}}_{t} & = \hbox{tanh}(\boldsymbol{W}_{c} [\boldsymbol{h}_{j-1}, \boldsymbol{x}_{j}] + \boldsymbol{b}_{c}) \\
    \boldsymbol{c}_{j} & = \boldsymbol{f}_{j} \circ \boldsymbol{c}_{j-1} + \boldsymbol{i}_{j} \circ \boldsymbol{\tilde{c}}_{j} \\
    \boldsymbol{o}_{j} & = \hat{\sigma}(\boldsymbol{W}_{o} [\boldsymbol{h}_{j-1}, \boldsymbol{x}_{j}] + \boldsymbol{b}_{o}) \\
    \boldsymbol{h}_{j} & = \boldsymbol{o}_{j} \circ \hbox{tanh}(\boldsymbol{c}_{j}).
\end{align}

Here $\boldsymbol{\tilde{c}}$ is the updated cell state, $\boldsymbol{c}_{j-1}$ and $\boldsymbol{h}_{j-1}$ represent the cell and hidden state from the previous time step, respectively. The input at the current time step is denoted by $\boldsymbol{x}_{j}$, $\boldsymbol{W}$ and $\boldsymbol{b}$ denote the weight matrices and bias vectors for the respective gates, and $\hat{\sigma}(\cdot)$ represents the sigmoid activation function. This formulation allows the network to dynamically regulate the flow of information across time steps.

To enhance the model's flexibility, after the LSTM processes the input sequence and generates the final state representing the time-ahead prediction, three dense layers are applied. This additional step produces the final forecast of the POD coefficients.

\begin{table}[h]
    \caption{Architecture details of the POD-DL model, where $k$ denotes the number of retained POD modes, $m$ represents the input window size, corresponding to the number of preceding time steps used to forecast the next one, and $B$ is the batch size.} \label{tab: pod_dl}
    \begin{tabular}{ccccc}
        \hline\hline
        \textbf{Layer} & \textbf{Name} & \textbf{\# Neurons} & \textbf{Activation function} & \textbf{Dimension}\\
        \hline
        $0$ & Input & - & - & $B \times m \times k$\\
        $1$ & LSTM & $100$ & ReLU & $B \times 1 \times 100$\\
        $2$ & Dense & $200$ & ReLU & $B \times 1 \times 200$ \\
        $3$ & Dense & $200$ & Linear & $B \times 1 \times 200$ \\
        $4$ & Dense & $k$ & Linear & $B \times 1 \times k$ \\
        \hline\hline
    \end{tabular}
\end{table}


Before training the DL model, the dataset composed of POD coefficients is standardized using the mean and standard deviation computed from the available data. Additionally, although the model operates autoregressively during inference, it is trained using a one-step-ahead prediction strategy, where it forecasts the next state in the sequence and compares it directly with the corresponding ground truth. This training setup is implemented using a rolling window technique as a preprocessing step.

This technique transforms a sequence of coefficients into structured input-output pairs suitable for training forecasting models. Given a dataset composed of temporally ordered coefficients, this approach constructs each training sample by selecting a fixed-length, $m$, sequence of consecutive coefficients (the window) as input, and the immediate subsequent coefficient as the target output. The window then shifts forward one time step, generating the next input-target pair, and the process repeats across the dataset. This method enables the model to learn temporal dependencies by framing the forecasting task as a supervised learning problem, where past states are used to predict future states.

The size of these windows depends on either the dynamics of the flow and the architecture of the DL model, as shown in \cite{2025_abadiaheredia_etal_poddl_ae}. Following this study, the window size set in this work is $m = 10$ for all testing cases defined in Sec. \ref{sec: results}.

Although the windows are generated sequentially, they are randomly shuffled prior to training in order to preserve statistical independence among training samples. Following standard training procedures, these windows are also reorganized into mini-batches to reduce the computational cost during training. The batch size, $B$, is usually set to be a multiple of $2$, in this work it is set as $B = 8$ for all testing cases defined in Sec. \ref{sec: results}.

To avoid exploding gradients, or collapsing, all weights are initialized following the Xavier \cite{xavier_initialization}, Kaiming \cite{kaiming_initialization} or orthogonal initialization, depending on their activation function and layer. This activation-dependent initialization is designed to preserve the variance of the input data across all layers of the model \cite{2016_goodfellow_etal_book}. For a detailed implementation of the initialization, we refer the reader to the publicly available code accompanying this work. The Adam method \cite{kingma2017adam} is used to optimize the training, with the default values for the parameters $\beta_{1} = 0.9$, $\beta_{2} = 0.999$ and $\epsilon = 10^{-8}$ as indicated in the original paper.

The loss function used to train this model is the regular mean squared error (MSE), defined as follows,

\begin{equation}
    \hbox{MSE} = \frac{1}{S}\sum_{i=0}^{S-1}(\boldsymbol{\gamma}^{i} - \boldsymbol{\hat{\gamma}}^{i})^{2},
    \label{eq: mse}
\end{equation}

where $\boldsymbol{\hat{\gamma}}^{i}$ is the $i$-th prediction obtained from the POD-DL model.

\begin{figure}[h]
	\centering
	\begin{subfigure}[b]{0.49\textwidth}
		\centering
		\includegraphics[width=\textwidth]{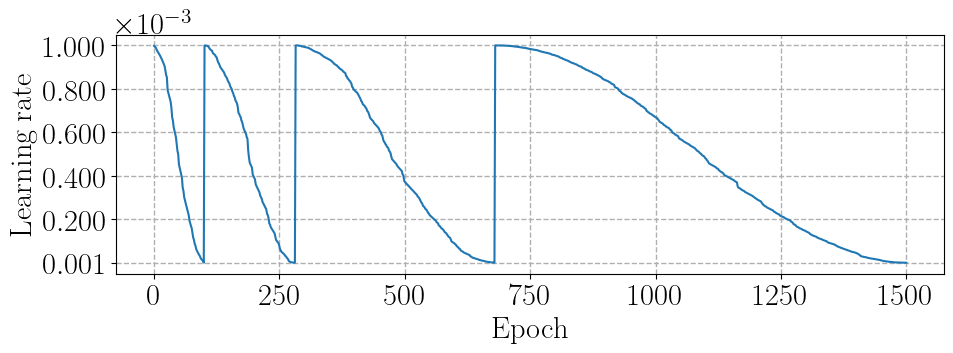}
        \caption{}
	\end{subfigure}
    \hfill
	\begin{subfigure}[b]{0.49\textwidth}
		\centering
		\includegraphics[width=\textwidth]{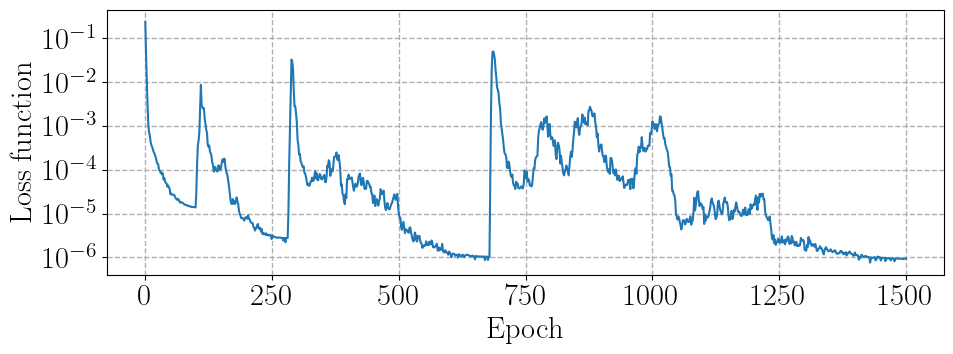}
        \caption{}
	\end{subfigure}
	\caption{Variation of the learning rate using the CosineAnnealingWarmRestarts scheduler \cite{lr_schedule} (left) and the corresponding loss function values (right) during a training process of 1500 epochs. The learning rate decays from a maximum value of $10^{-3}$ to a minimum value of $10^{-6}$.}
	\label{fig: lr_schedule}
\end{figure}

A learning rate scheduler known as CosineAnnealingWarmRestarts \cite{lr_schedule}, was employed to prevent the loss function from becoming trapped in local minima. This scheduler operates by gradually decaying the learning rate from an initial maximum value to a specified minimum over a certain number of epochs (less than the total training length). Following this decay, the learning rate resets to the maximum value and begins a new decay phase, which can either span the same or double, as in this work, the number of epochs as the previous phase. This cyclic process repeats throughout the training period. Figure \ref{fig: lr_schedule} illustrates the learning rate decay during training alongside the corresponding loss function values.

The following section provides a detailed description of the implementation of the adaptive framework, presented in Sec. \ref{sec: adaptive_framework}, including the key hyperparameters that influence its performance, as well as the metric employed to quantify the accuracy of predictions.

\section{Implementation of the adaptive framework}\label{sec: implementation_adaptive}

The adaptive framework described in Sec. \ref{sec: adaptive_framework} involves several hyperparameters that condition both the results obtained and how fast the DL model is trained. Among these, a subset has been identified as particularly impactful to the overall framework performance and listed in Tab. \ref{tab: model_hyperparameters}.

\begin{table}[h]
    \centering
    \caption{Hyperparameters defining the behavior of the adaptive framework. A short description of the hyperparameter is found in left column, and a the symbol used to identify them is defined in the right column.\label{tab: model_hyperparameters}}
    
    \begin{tabular}{lc}
        \hline\hline
        \textbf{Hyperparameter} & \textbf{Symbol} \\
        \hline
        Training length (Epochs) & $E$ \\
        Number snapshots for the first training & $S_{0}$ \\
        Number of snapshots for subsequent trainings & $S_{1}$ \\
        Number of predictions & $P$ \\
        \hline\hline
    \end{tabular}
    
\end{table}

\begin{figure}[h]
    \centering
    \includegraphics[width=\textwidth]{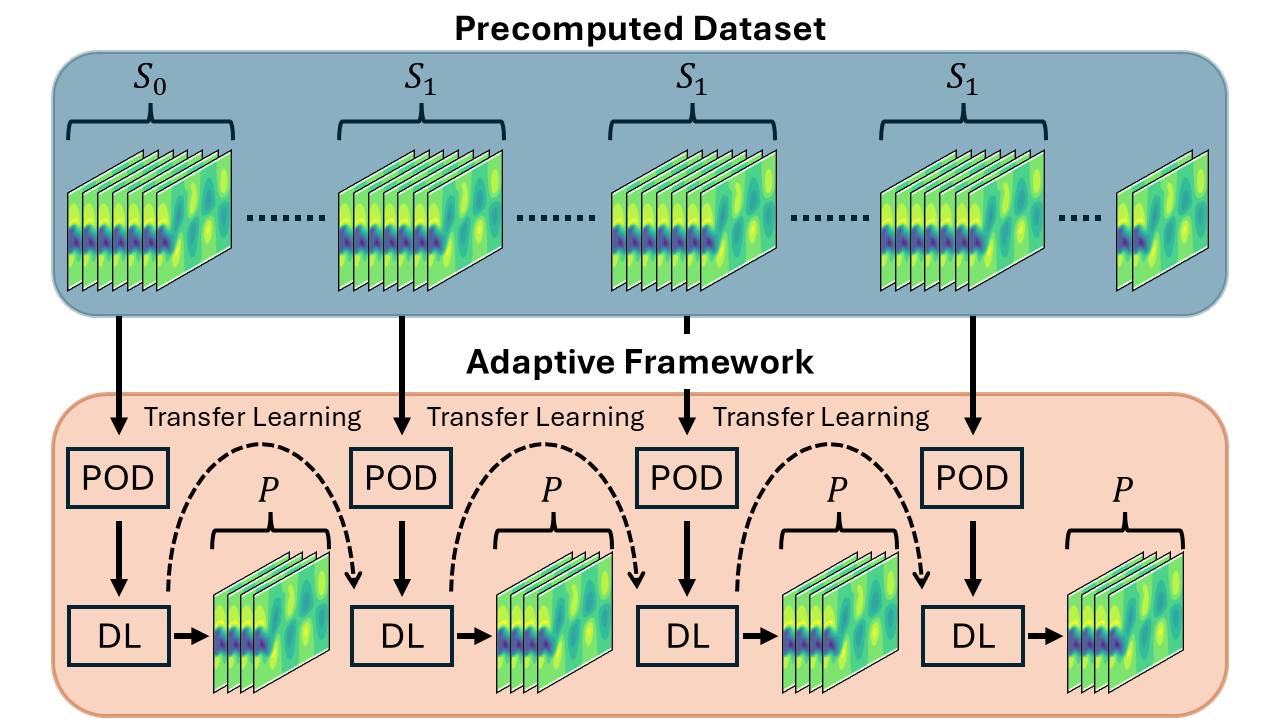}
    \caption{Visual representation of the adaptive framework proposed in this work. Initially, a set of $S_{0}$ snapshots is extracted from a precomputed dataset to train the POD-DL model. Within this model, POD is applied to obtain the corresponding POD coefficients, which are then used to train a deep learning model, for $E$ epochs. This trained model forecasts $P$ future snapshots, after which an additional set of $S_{1}$ snapshots is drawn from the dataset to retrain the POD-DL model, with the same number of epochs. During each retraining step, the DL model leverages transfer learning by initializing its weights with those obtained from the previous training phase. This adaptive procedure is repeated iteratively until predictions cover the entire temporal domain of interest.}
	\label{fig: adaptive_framework}
\end{figure}

These hyperparameters were identified after extensive experimentation, where it was determined that the most critical to the performance of the adaptive framework are the number of snapshots used for the initial training, $S_{0}$, and subsequent retrainings, $S_{1}$, as well as the number of predictions, $P$. Variations in the training length, $E$, did not result in significant changes in the model's performance. A visual depiction of the adatptive framework proposed in this work is shown in Fig. \ref{fig: adaptive_framework}.

From Sec. \ref{sec: adaptive_framework}, the adaptive framework begins with an initial set of snapshots, drawn from the precomputed dataset, to perform the first training of the POD-DL model described in Sec. \ref{sec: dl_model}. The number of initial snapshots, $S_{0}$, depends on the specific problem and, more notably, on the complexity of the flow dynamics. As demonstrated below, for laminar flows, $S_{0}$ can be smaller than that required for turbulent flows. This difference arises from the increased complexity of turbulent dynamics, which typically necessitates a larger number of snapshots for effective training.

If the combined number of training snapshots and predicted snapshots does not cover the entire temporal domain of interest, additional data, $S_{1}$, is then drawn from the precomputed dataset to retrain the POD-DL model. This number, $S_{1}$, remains fixed for all subsequent retraining iterations. 

The retraining is only done over the new $S_{1}$ snapshots, discarding the previous ones. To alleviate the problem of using a reduced dataset of $S_{1}$ samples, the adaptive framework leverages a transfer learning procedure, wherein the POD-DL model initializes each subsequent training session with the optimal weights obtained from the previous training phase, ensuring continuity and efficiency in the learning process. The benefits of transfer learning in the POD-DL model was already observed in \cite{2025_abadiaheredia_etal_poddl_ae}.

Lastly, the number of training epochs, $E$, is determined by both the dynamics of the system and the learning rate scheduler described in Sec. \ref{sec: dl_model}, where it is preferable for the training process to conclude at the end of a decay cycle, allowing the learning rate to reach its minimum value as shown in Fig. \ref{fig: lr_schedule}, thereby optimizing the training performance.

In this work, after training, the POD-DL model consistently generates a fixed number of predictions, $P$. The implementation of an uncertainty quantification method to assess the reliability of these predictions is deferred to future research. However, various approaches could be explored for this purpose, including regular bootstrap techniques \cite{uq_adaled, adaled_petros}, methods leveraging the structure of POD \cite{rapun_adaptive_pod_modes_2015, beltran_adaptive, uq_abadiaheredia_etal}, conformal prediction frameworks \cite{uq_conformal_prediction}, or even by evaluating the residuals of the governing PDEs \cite{rapun_adaptive_pod_modes_2015}.

To quantify the reliability of predictions from POD-DL, in this work it is used a metric identify as the probability of prediction error (PPE). This metric represents the probability that, upon randomly selecting a coordinate from the predicted snapshot, the absolute prediction error will be less than or equal to an error threshold, $\alpha > 0$. It is mathematically defined as follows,

\begin{equation} \label{eq: ppe}
    \text{Pr}_{\alpha}(t_{S}) \equiv \text{Pr}(-\alpha \leq \boldsymbol{v}_{s} - \boldsymbol{\hat{v}}_{s} \leq \alpha).
\end{equation}

Here $\boldsymbol{v}_{s}$ and $\boldsymbol{\hat{v}}_{s}$ denote the $s$-th ground truth and predicted snapshots, respectively. It is important to note that the error threshold ($\alpha$) must be selected based on the specific characteristics of the problem under study. The pseudocode used to compute this probability is provided in \ref{appnx: pape}.

Note that a high PPE value for a small threshold error, relative to the dynamics of the problem, indicates greater predictive accuracy. Therefore, it is always desirable for this probability to be as high as possible ($\text{Pr}_{\alpha}(t_{S}) \simeq 1$).

Furthermore, since the aim of this work is to propose an adaptive framework that reduces the computational cost of solving a dynamical system relative to traditional numerical solvers, the following metric is introduced to approximate the theoretical computational saving (TCS) achieved.

\begin{equation} \label{eq: tcs}
    \text{TCS} = (1 - (S_{0} + n S_{1}) / S) \times 100\%.
\end{equation}

Here, $n$ represents all the times additional data was drawn from the precomputed dataset, and $S$ denotes the total number of snapshots in the dataset, as shown in eq. \eqref{eq: dataset}. This metric provides an estimate of the percentage reduction in computational cost, reflecting the savings achieved by relying on the adaptive framework instead of computing all snapshots via a traditional numerical solver.

This adaptive framework was tested on three different datasets describing the dynamics of different flows at different regimes. The following section show the results for each case obtained at different values of $S_{0}$, $S_{1}$ and $P$.

\section{Results}\label{sec: results}

As stated in section \ref{sec: adaptive_framework} the adaptive framework studied in this work considers an already computed dataset, from where to obtain highly accurate data to train the POD-DL model. Specifically, three distinct datasets are employed, each representing a different flow regime: the three-dimensional wake of a circular cylinder in both laminar and turbulent conditions, and an isothermal subsonic jet under turbulent conditions.

These flows are governed by the Navier–Stokes equations, expressed in their non-conservative form as follows:

\begin{align}
    \nabla \cdot \boldsymbol{v} & = 0, \\
    \frac{\partial \boldsymbol{v}}{\partial t} + (\boldsymbol{v} \cdot \nabla)\boldsymbol{v} & = -\nabla \boldsymbol{p} + \frac{1}{\hbox{Re}}\Delta \boldsymbol{v},
\end{align}

where $\boldsymbol{p}$ and $\boldsymbol{v}$ represent the pressure field and the velocity field, respectively, and Re is the Reynolds number, defined as $\hbox{Re} = VL/\nu$. Here, $V$ denotes the free stream velocity, $L$ the characteristic length and $\nu$ the kinematic viscosity of the fluid.

For the remainder of this work, and for the sake of simplicity, the datasets representing the three-dimensional wake of a circular cylinder, in laminar and turbulent regimes, will be referred to as the laminar flow and turbulent flow, respectively.

\subsection{Three-dimensional wake of a circular cylinder at laminar regime: laminar flow} \label{sec: results_laminar_cyl}

The first dataset represents the three-dimensional wake of a circular cylinder in a viscous, incompressible and Newtonian flow, at Re = $280$. This dataset was generated and validated by Le Clainche {\it et al.} \cite{LeClainche_2018_cilind}, where we refer for more information about the simulation.

This dataset $\boldsymbol{\mathring{D}}$ is composed by $499$ snapshots covering the entire temporal domain $\mathcal{P}_{t}$, such that $\boldsymbol{\mathring{D}} \in \mathbb{R}^{2 \times 100 \times 40 \times 64 \times 499}$. It also describes $C = 2$ velocity components, in the streamwise and wall-normal directions, respectively.

In this particular case, the performance of the POD-DL model exhibits limited sensitivity to the choice of hyperparameter values, attributable to the simplicity of the underlying dynamics. Three experiments were conducted on this dataset, varying the hyperparameters $S_{1}$ and $P$. The value of $S_{0}$ was intentionally selected to be relatively small in comparison to the total number of available snapshots, for this dataset, it was set to  $S_{0} = 100$. The specific values of all hyperparameters used in these tests are summarized in Table \ref{tab: cyl_laminar_hyperparameters_cases}.

\begin{table}[h]
    \centering
    \caption{Hyperparameters used to test the adaptive framework on the laminar flow around a circular cylinder. In the second test the number prediction $P$ is increased after the second training, from $P = 100$ to $P = 200$.} \label{tab: cyl_laminar_hyperparameters_cases}
    
    \begin{tabular}{lccccc}
        \hline\hline
        \textbf{Test} & \textbf{$P$} & \textbf{$S_{0}$} & \textbf{$S_{1}$} & \textbf{$E$}\\
        \hline
        T1 & $100$ & $100$ & $50$ & $1500$ \\
        T2 & $100$/$200$ & $100$ & $100$ & $1500$ \\
        T3 & $400$ & $100$ & $50$ & $1500$ \\
        \hline\hline
    \end{tabular}

\end{table}

The adaptive framework operates on the velocity field, where POD is applied to decompose this field into its corresponding POD modes and coefficients, as described in eq. \eqref{eq: svd}. The DL model is then trained on the POD coefficients of this field. For the initial training with $S_{0}$ snapshots, it is necessary to determine the number of POD modes to retain for forecasting. This decision is typically based on the percentage of energy captured within the dynamics for various truncation levels, represented by the cumulative energy, which is defined as follows,

\begin{equation}
    \hbox{Eg}(k) = \frac{\sum^{k}_{i=1}\sigma_{i}}{\sum^{S_{0}}_{i = 1} \sigma_{i}}. \label{eq: cummEnerg}
\end{equation}

\begin{figure}[H]
    \centering
    \includegraphics[width=0.4\textwidth]{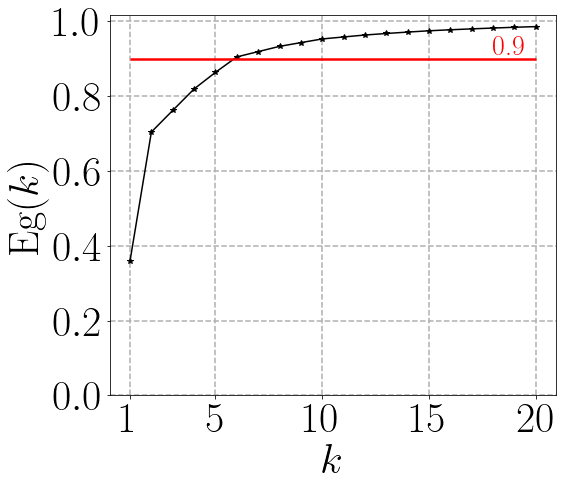}
    \caption{Percentage of energy contained, Eg$(k)$, when the truncation is done at the $k$, up to $20$, most energetic modes in the laminar flow past a cylinder.}
    \label{fig: cil_laminar_sing_val}
\end{figure}

Here $k$ denotes the truncation level, representing the number of retained POD modes, and $\sigma_{i}$ is the $i$-th singular value. Figure \ref{fig: cil_laminar_sing_val} illustrates the cumulative energy, calculated by applying POD to the first $S_{0}$ snapshots, as a function of the truncation level for the laminar flow past a cylinder. For flows of this type, it is common to retain $90\%$ of the total energy within the dynamics, which, in this case, corresponds to keeping the first $6$ POD modes.

This truncation is kept for all subsequent trainings within the adaptive framework. For this reason, it is required that $S_{1} \geq k$; otherwise, the number of POD modes computed from the new $S_{1}$ snapshots would be insufficient to match the selected truncation level $k$.

The reliability of predictions will be assessed using the PPE metric, defined in eq. \eqref{eq: ppe}, along with percentil-based comparisons. Importantly, this evaluation will be conducted not on the velocity field directly, but on the vorticity field, which is derived from the velocity field as follows:

\begin{equation} \label{eq: vorticity}
    \boldsymbol{\omega} = \nabla \times \boldsymbol{v} = \left( \frac{\partial \boldsymbol{v}_{y}}{\partial x} - \frac{\partial \boldsymbol{v}_{x}}{\partial y}\right).
\end{equation}

This choice is motivated by the fact that the vorticity field provides a more intuitive representation of the underlying flow physics. For instance, it facilitates the evaluation of how accurately the POD-DL model captures the position and structure of vortical features, by allowing direct comparison with the reference solution obtained from the numerical simulation. 

\begin{figure}[H]
	\centering
	\begin{subfigure}[b]{0.3\textwidth}
		\centering
		\includegraphics[width=\textwidth]{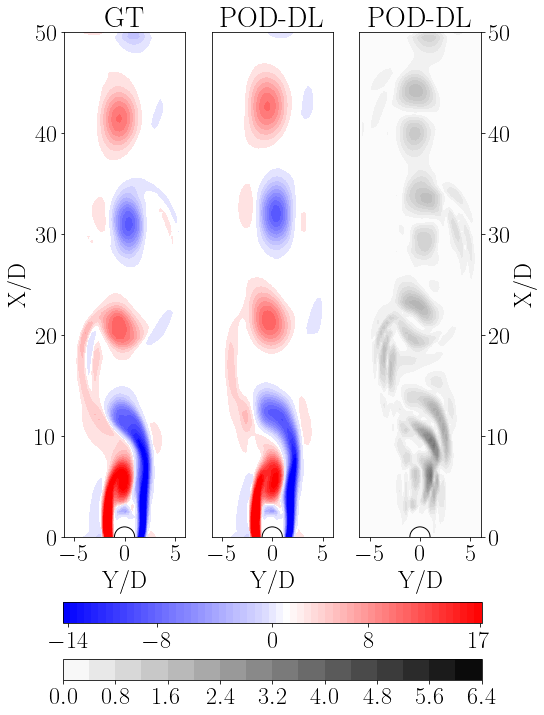}
        \caption{$t_{S} = 300$}
	\end{subfigure}
    \hfill
	\begin{subfigure}[b]{0.3\textwidth}
		\centering
		\includegraphics[width=\textwidth]{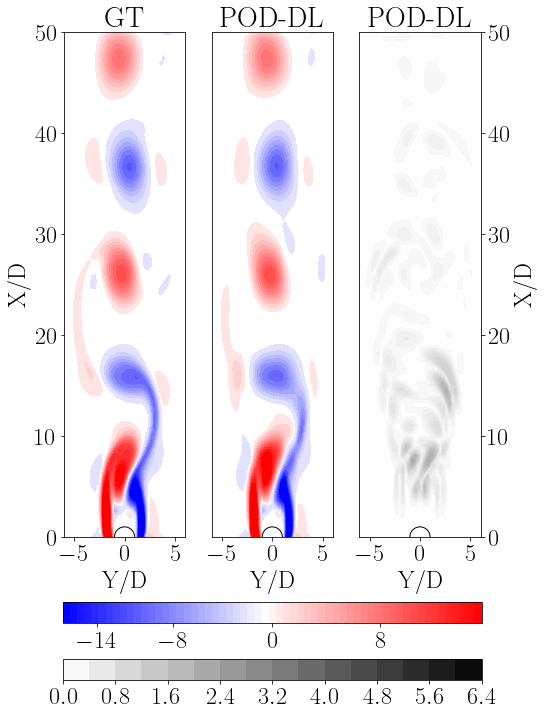}
        \caption{$t_{S} = 150$}
	\end{subfigure}
    \hfill
	\begin{subfigure}[b]{0.3\textwidth}
		\centering
		\includegraphics[width=\textwidth]{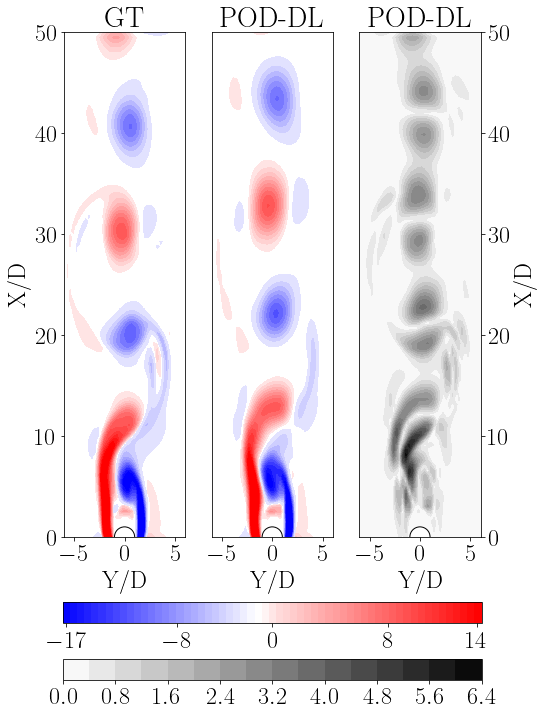}
        \caption{$t_{S} = 450$}
	\end{subfigure}
	\caption{Snapshots comparing the ground truth (GT) vorticity, of the laminar flow past a cylinder, with the one predicted from POD-DL, at some representative time steps for test cases T1 (a), T2 (b) and T3 (c), respectively. Here $t_{S}$ indicates the snapshot index rather than the actual time $t$.}
	\label{fig: cil_laminar_snap}
\end{figure}

Accordingly, Fig. \ref{fig: cil_laminar_snap} presents a visual comparison between the predicted and actual vorticity fields at selected time steps, indicating that the vorticity values predominantly lie within the absolute interval [$7$, $16$]. Based on this range, an absolute prediction error of $1$, in the vorticity, corresponds to a relative error interval of $[1 / 16, 1 / 7] \times 100\% = [6.25\%, 14.2\%]$. Similarly, an absolute error of $0.5$ corresponds to a relative error range of $[3.12\%, 7.14\%]$. To ensure that the POD-DL predictions remain close to the reference dynamics, it is desirable to achieve high probabilities of obtaining absolute prediction errors within these thresholds.

\begin{figure}[H]
	\centering
	\begin{subfigure}[b]{0.49\textwidth}
		\centering
		\includegraphics[width=\textwidth]{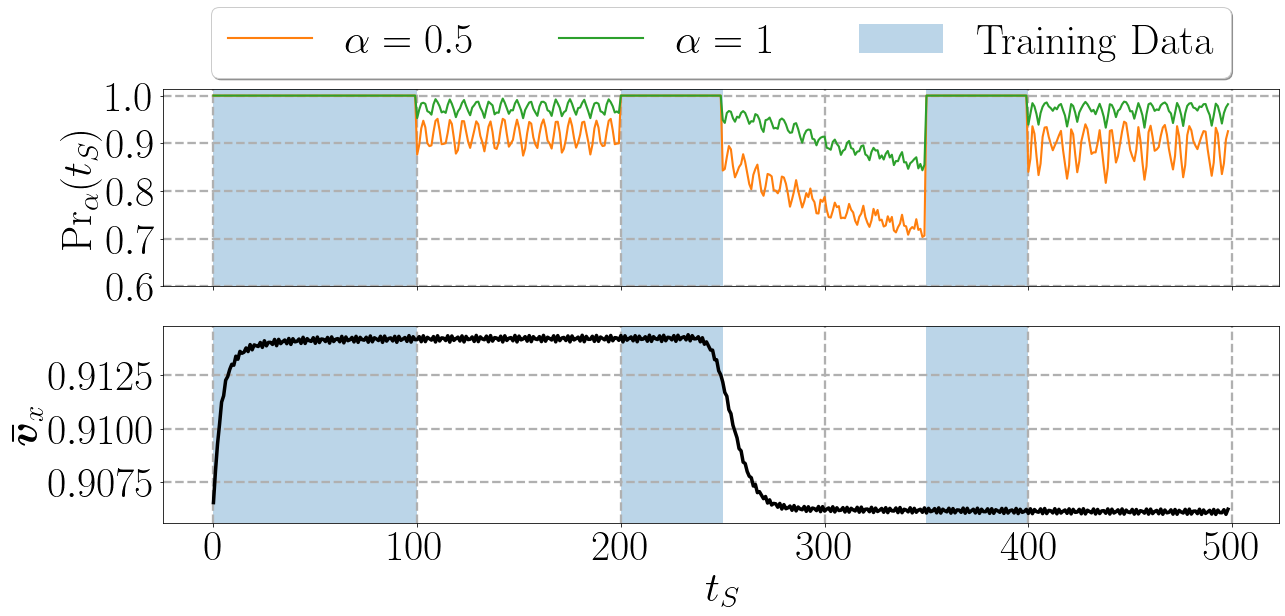}
        \caption{}
	\end{subfigure}
    \hfill
	\begin{subfigure}[b]{0.49\textwidth}
		\centering
		\includegraphics[width=\textwidth]{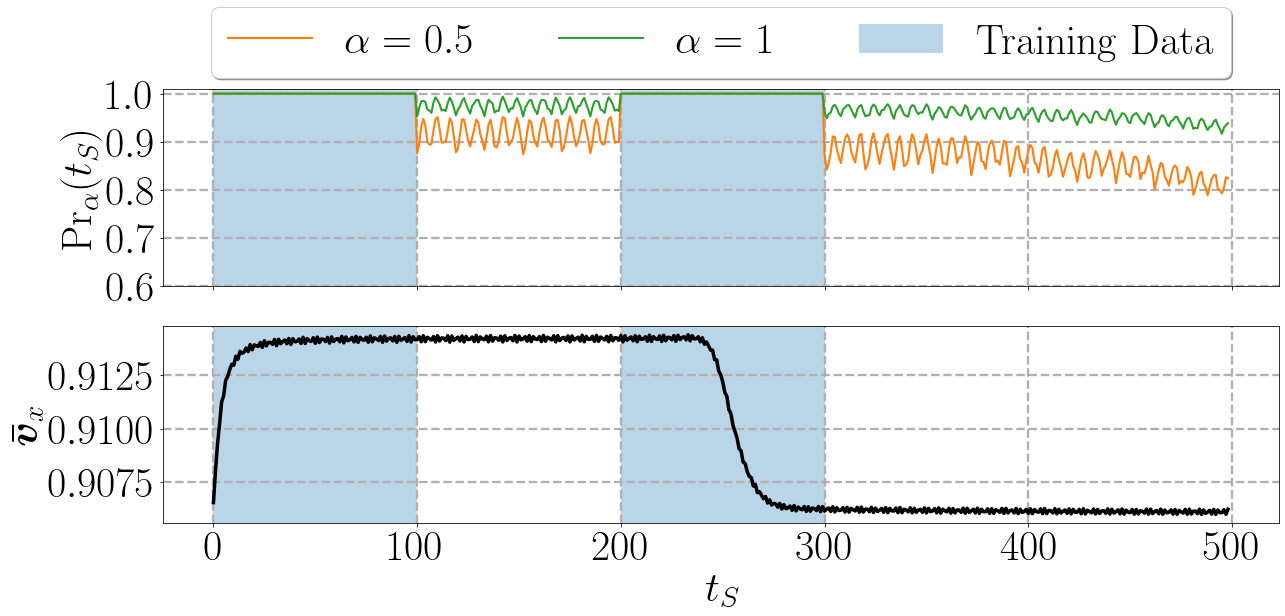}
        \caption{}
	\end{subfigure}
    \hfill
	\begin{subfigure}[b]{0.49\textwidth}
		\centering
		\includegraphics[width=\textwidth]{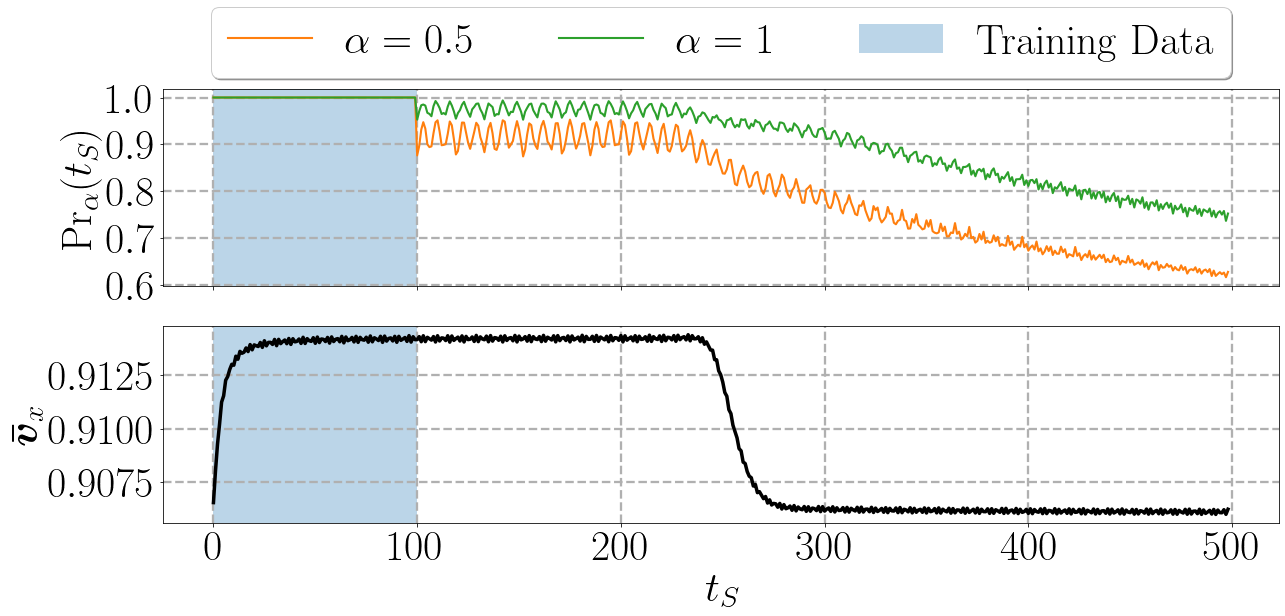}
        \caption{}
	\end{subfigure}
	\caption{The top row of all figures displays the probability of achieving an absolute vorticity prediction error below thresholds of 0.5 (orange) and 1 (green) for the three test cases: T1 (a), T2 (b), and T3 (c), in the laminar flow. The bottom row presents the temporal evolution of the spatially averaged streamwise velocity. These error thresholds are selected based on the typical range of vorticity values identified in the flow, as evident in Fig. \ref{fig: cil_laminar_snap}.}
	\label{fig: cil_prob_err}
\end{figure}

Therefore, Fig. \ref{fig: cil_prob_err} presents the probability of achieving an absolute vorticity prediction error less than or equal to $0.5$ and $1$, respectively, across the three test cases described in Tab. \ref{tab: cyl_laminar_hyperparameters_cases}. Specifically, this is measured by the PPE metric, and represented by the probability $\text{Pr}_{\alpha}(t_{S}) \equiv \text{Pr}(-\alpha \leq \boldsymbol{\omega}_{s} - \boldsymbol{\hat{\omega}}_{s} \leq \alpha)$ for $\alpha \in \{0.5, 1\}$, where $\boldsymbol{\omega}_{s}$ and $\boldsymbol{\hat{\omega}}_{s}$ denote the $s$-th ground truth and predicted vorticity snapshots, respectively. Note that the snapshots for which the probability reaches $100\%$ ($Pr_{\alpha}(t_{S}) = 1$) correspond to those used for training the POD-DL model.

This figure also presents the evolution of the spatially averaged streamwise velocity, providing a visual reference for the flow dynamics captured by the snapshots used in training the model, both during the initial and subsequent retraining phases.

Note in Figs. \ref{fig: cil_prob_err} (a) and (c), which correspond to test cases T1 and T3, an interesting effect associated with the distributional shift, around snapshot $250$, in the flow dynamics can be observed. In both cases, the probability of achieving low prediction errors begins to decline precisely at the moment when the dynamics undergo a distribution change, as highlighted in the bottom row of each figure. In both cases, this indicates that predictions have begun to diverge from actual dynamics.

However, as shown in Fig. \ref{fig: cil_prob_err} (b) for test case T2, this issue is mitigated through the inclusion of new training data that capture the distributional shift, allowing the POD-DL model to account for it during retraining. This enables the subsequent predictions to adjust to the change in the dynamics.

Even in test case T1, where a divergence between the predictions and the actual dynamics is observed precisely at the point of distributional change, this divergence is halted upon the introduction of new data for retraining, even with a limited number of samples. This highlights the advantages of transfer learning within the adaptive framework, enabling the model to effectively leverage prior training to adapt to newly emerging dynamics.

\begin{figure}[H]
	\centering
	\begin{subfigure}[b]{0.49\textwidth}
		\centering
		\includegraphics[width=\textwidth]{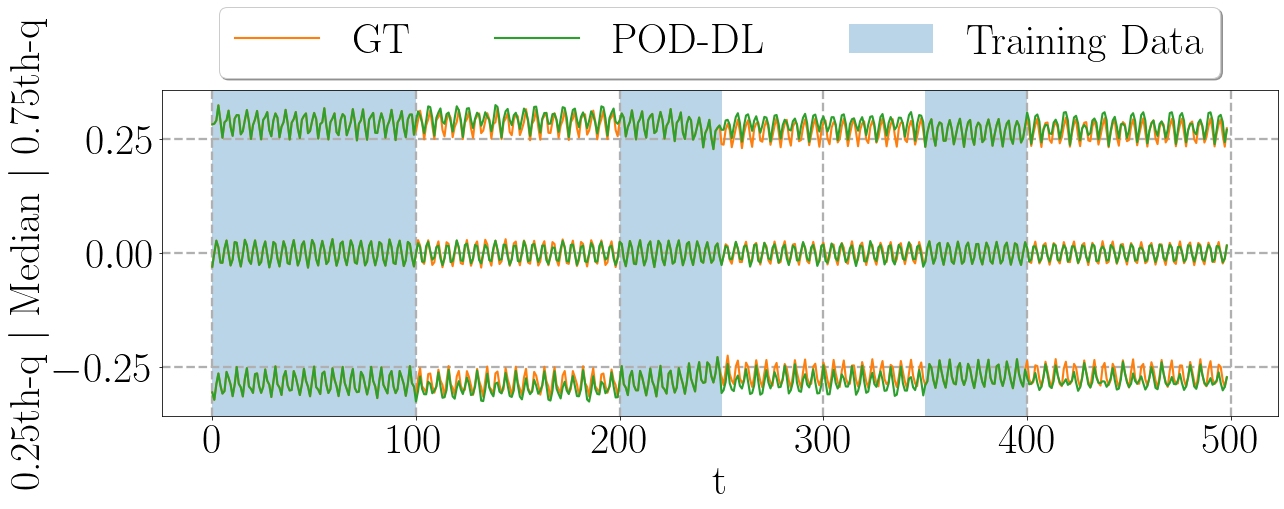}
        \caption{}
	\end{subfigure}
    \hfill
	\begin{subfigure}[b]{0.49\textwidth}
		\centering
		\includegraphics[width=\textwidth]{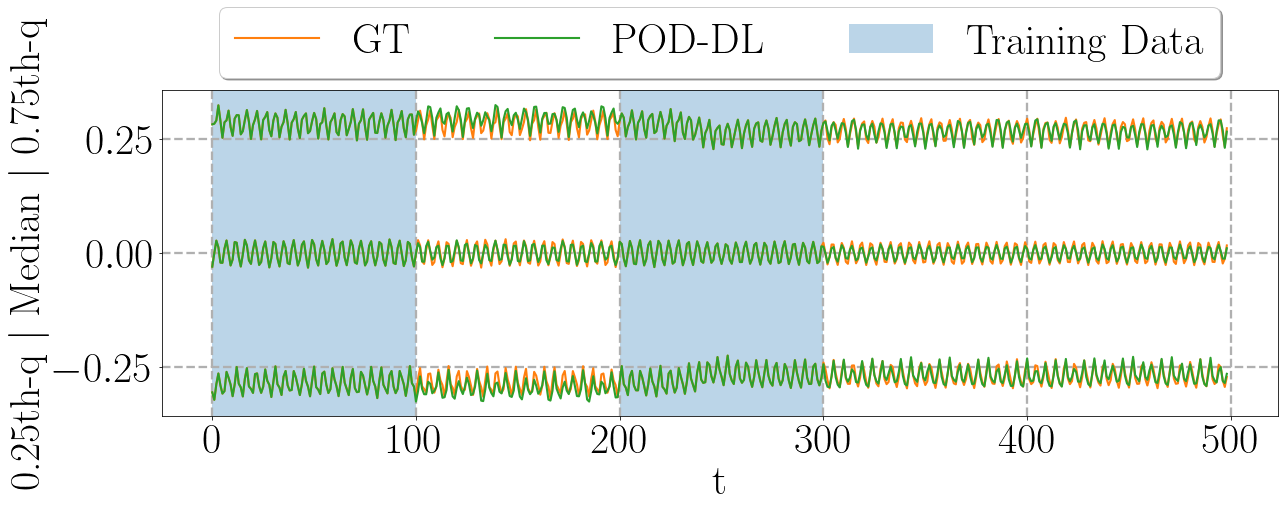}
        \caption{}
	\end{subfigure}
    \hfill
	\begin{subfigure}[b]{0.49\textwidth}
		\centering
		\includegraphics[width=\textwidth]{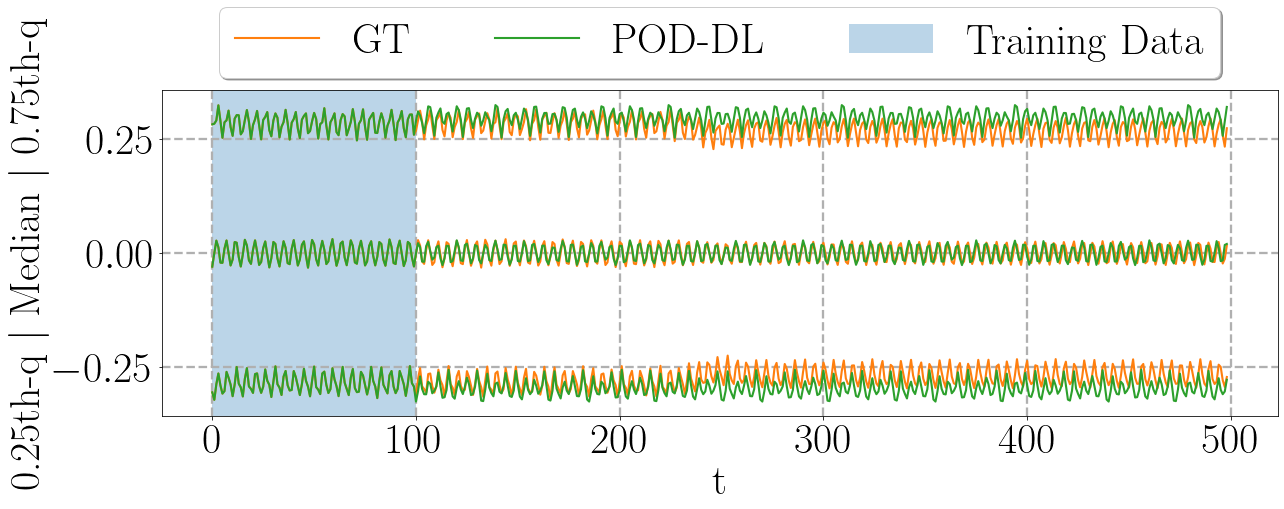}
        \caption{}
	\end{subfigure}
	\caption{Temporal evolution of the 25th (bottom), 50th (middle), and 75th (top) vorticity percentiles from ground truth (blue) and POD-DL predictions (orange) for test cases 1 (a), 2 (b) and 3 (c) in the laminar flow scenario. Shaded regions indicate training snapshots used in both the initial and retraining phases.}
	\label{fig: cil_quantiles}
\end{figure}

Figure \ref{fig: cil_quantiles} shows a comparison of the $0.25$-th, $0.5$-th (median) and $0.75$-th percentiles computed at every snapshot of the ground truth vorticity and the one obtained from the adaptive framework. From this figure, we observe the same phenomenon described earlier: predictions for test cases T1 and T3 begin to diverge starting from snapshot $250$, coinciding with a distribution shift in the dynamics. In contrast, this divergence does not occur in test case T2, where the newly introduced training data captures the distributional change, enabling the POD-DL model to adjust accordingly. Furthermore, in test case T1, although the predictions initially deviate, they appear to converge back to the actual dynamics once the model is retrained using data reflective of the new dynamics.

After extensive experimentation, it was observed that training the POD-DL model with $6$ POD coefficients and a training set of $100$ samples takes approximately $1.2$ minutes on a CPU Intel(R) Core(TM) i7-10700 CPU @ 2.90GHz with $16$ cores and $64$ Gb of RAM. The computation time for predictions, regardless of the number of predictions, is nearly negligible, averaging around $2$ seconds. In this context, can be assumed that the theoretical computational saving achieved, measured by TCS in eq. \eqref{eq: tcs}, for test cases T1 and T2, where $299$ out-of $499$ time steps were predicted, is approximately $60\%$, while the theoretical saving for test case $T3$, with $399$ predicted time steps, reaches $80\%$, being this last one the less accurate. However, these results demonstrate the effectiveness of this adaptive framework for laminar flows.

\subsection{Three-dimensional wake of a circular cylinder at turbulent regime: turbulent flow}\label{sec: results_turbulent_cyl}

Similar to the laminar flow case discussed in Section \ref{sec: results_laminar_cyl}, the dataset analyzed in this section also represents the wake of a circular cylinder. The key difference, however, is that it corresponds to a turbulent regime, with a Reynolds number of Re = $4000$.

This dataset $\boldsymbol{\mathring{D}}$ consists of a collection of $2000$ snapshots defining the two-dimensional velocity field, $\boldsymbol{v} = (\boldsymbol{v}_{x}, \boldsymbol{v}_{y})$, such that $\boldsymbol{\mathring{D}} \in \mathbb{R}^{2 \times 301 \times 111 \times 1 \times 2000}$. It has been obtained from an experimental analysis, performed by \cite{Mendez_2020_experimental}, of a three-dimensional wake behind a circular cylinder, with a diameter of $D = 5$. These experiments were conducted in the L10 low-speed wind tunnel at the von Karman Institute, employing time-resolved particle image velocimetry (TR-PIV) for data acquisition.

This dataset corresponds only to the initial steady state of the flow at Re = $4000$, which subsequently transitions to another steady state at Re = $2600$, passing through a transient state during the process. These last two states are not represented in the dataset.

Several variations of the hyperparameters within the adaptive framework were explored to study the performance of the framework to highly complex dynamics. These variations are summarized in Table \ref{tab: cyl_vki_hyperparameters_cases}, where the amount of data used for the initial training was again deliberately chosen to be relatively small compared to the total number of snapshots, with $S_{0} = 200$.

\begin{table}[h]
    \centering
    \caption{Hyperparameters used to test the adaptive framework on the turbulent flow past a circular cylinder.\label{tab: cyl_vki_hyperparameters_cases}}
    
    \begin{tabular}{lccccc}
        \hline\hline
        \textbf{Test} & \textbf{$P$} & \textbf{$S_{0}$} & \textbf{$S_{1}$} & \textbf{$E$}\\
        \hline
        $T1$ & $50$ & $200$ & $100$ & $1500$ \\
        $T2$ & $100$ & $200$ & $100$ & $1500$ \\
        $T3$ & $100$ & $200$ & $200$ & $1500$ \\
        $T4$ & $500$ & $200$ & $200$ & $1500$ \\
        \hline\hline
    \end{tabular}

\end{table}

\begin{figure}[H]
    \centering
    \includegraphics[width=0.4\textwidth]{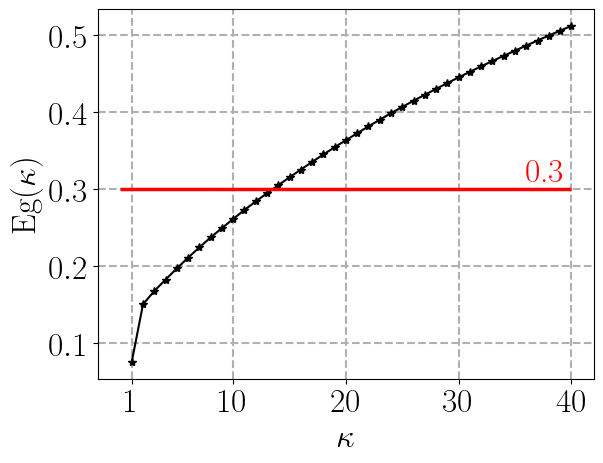}
    \caption{Percentage of energy contained, Eg$(k)$, when the truncation is done at the $k$, up to $40$, most energetic modes in the turbulent flow past a cylinder.}
    \label{fig: cil_turbulent_sing_val}
\end{figure}

Similar to the laminar flow, the POD is applied to the first $S_{0} = 200$ snapshots to compute the corresponding POD coefficients. The truncation of POD modes is then determined by the amount of energy to be retained within the dynamics, using the cummulative energy function, eq. \eqref{eq: cummEnerg}. In similar studies \cite{pod_40_energy_causality}, it has been shown that for turbulent flows, retaining around $30\%$ of the total energy is sufficient to capture the principal dynamics of interest. As illustrated in Fig. \ref{fig: cil_turbulent_sing_val}, this energy threshold is achieved by retaining the first $13$ modes.

This truncation is maintained for all training iterations in the adaptive framework to maintain consistency with the POD-DL model architecture, which remains intact throughout the adaptive process. Similar to the laminar flow, the POD-DL model is trained on snapshots from the velocity field, while testing of the predictions is performed on the vorticity field, computed as indicated in eq. \eqref{eq: vorticity}.

For the turbulent flow, the accuracy of predictions in approximating the actual flow dynamics is evaluated using the PPE metric, defined in eq. \eqref{eq: ppe}, along with comparisons based on percentiles and turbulent kinetic energy (TKE). The TKE is defined as one-half the sum of the variances of the fluctuating components of the velocity field. Mathematically, it can be expressed as:
\begin{align}
    \text{TKE} & = \frac{1}{2} \left(\overline{(\boldsymbol{\tilde{v}}_{x})^{2}} + \overline{(\boldsymbol{\tilde{v}}_{y})^{2}}\right), \label{eq: tke} \\
    \overline{(\boldsymbol{\tilde{v}}_{x})^{2}} & = \frac{1}{T} \int_{0}^{T} \boldsymbol{\tilde{v}}_{x}(t)^{2} dt = \frac{1}{T} \int_{0}^{T} (\boldsymbol{v}_{x}(t) - \boldsymbol{\bar{v}}_{x})^{2} dt, \\
    \boldsymbol{\bar{v}}_{x} & = \frac{1}{T} \int_{0}^{T} \boldsymbol{v}_{x}(t) dt.
\end{align}

Here, is $\boldsymbol{\tilde{v}}_{x}$ represents the fluctuations in the streamwise velocity, obtained by subtracting the time-averaged mean flow from the instantaneous velocity, i.e., $\boldsymbol{\tilde{v}}_{x}(t) = \boldsymbol{v}_{x}(t) - \boldsymbol{\bar{v}}_{x}$. The same procedure is applied to the wall-normal velocity component, $\boldsymbol{v}_{y}$, to isolate its fluctuating part from the mean flow. The TKE is employed as a diagnostic to assess how accurately the dynamics predicted by the adaptive framework adhere to the underlying physics of the flow.

\begin{figure}[H]
	\centering
	\begin{subfigure}[b]{0.32\textwidth}
		\centering
		\includegraphics[width=\textwidth]{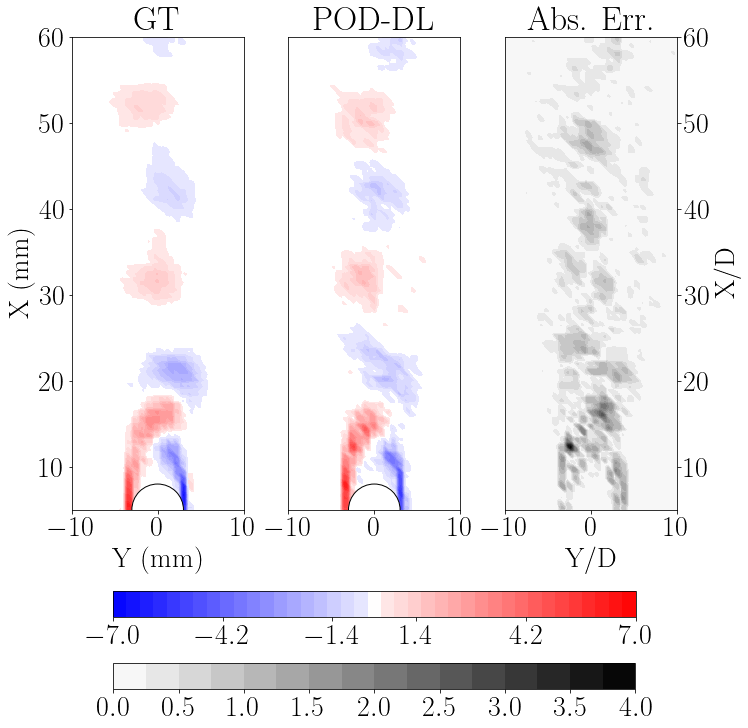}
        \caption{$t_{S} = 150$}
	\end{subfigure}
    \hfill
	\begin{subfigure}[b]{0.32\textwidth}
		\centering
		\includegraphics[width=\textwidth]{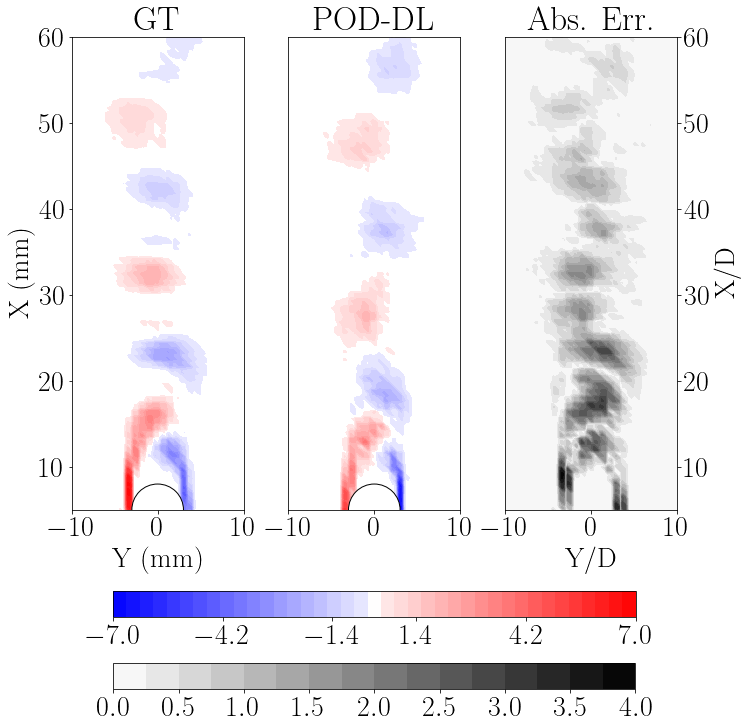}
        \caption{$t_{S} = 325$}
	\end{subfigure}
    \hfill
	\begin{subfigure}[b]{0.32\textwidth}
		\centering
		\includegraphics[width=\textwidth]{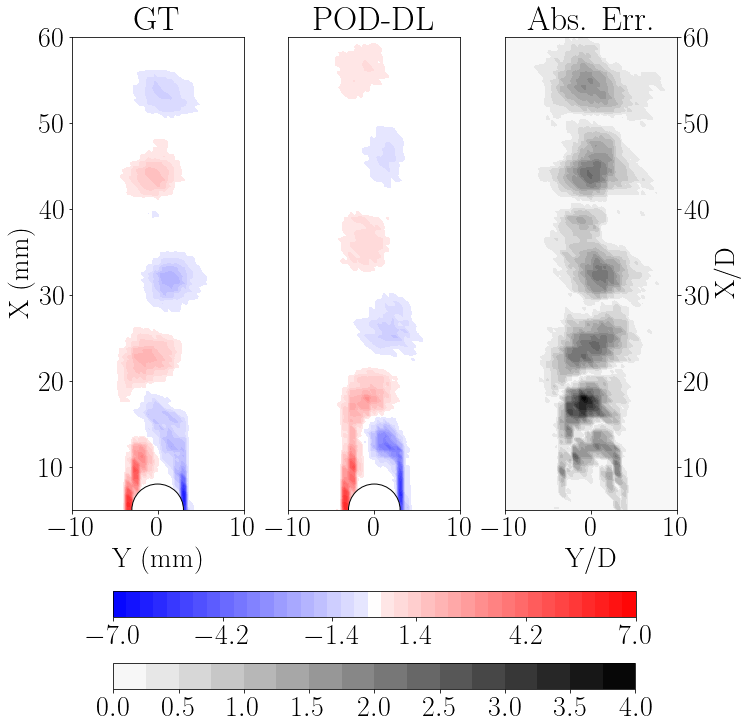}
        \caption{$t_{S} = 375$}
	\end{subfigure}
	\caption{Snapshots comparing the ground truth (GT) vorticity, of the turbulent flow past a cylinder, with the one predicted from POD-DL, at some representative time steps for test cases T1 (a), T3 (b) and T4 (c), respectively. Here $t_{S}$ indicates the snapshot index rather than the actual time $t$.}
	\label{fig: vki_snap}
\end{figure}

Similar to the laminar flow, Fig. \ref{fig: vki_snap} presents a visual comparison between the predicted and actual vorticity fields at selected time steps, indicating that the vorticity values predominantly lie within the absolute interval [$2.5$, $7$], which is narrower than the interval identified in the laminar flow scenario. Based on this range, an absolute prediction error of $0.5$, in the vorticity, corresponds to a relative error interval of $[0.5 / 7, 0.5 / 2.5] \times 100\% = [7.1\%, 20\%]$. Similarly, an absolute error of $0.2$ corresponds to a relative error range of $[2.8\%, 8\%]$. These error magnitudes can be considered small for this flow at turbulent regime, particularly given that the POD-DL model operates on flow structures with coherent behavior, since only the most energetic modes were retained for prediction.

\begin{figure}[H]
	\centering
	\begin{subfigure}[b]{0.49\textwidth}
		\centering
		\includegraphics[width=\textwidth]{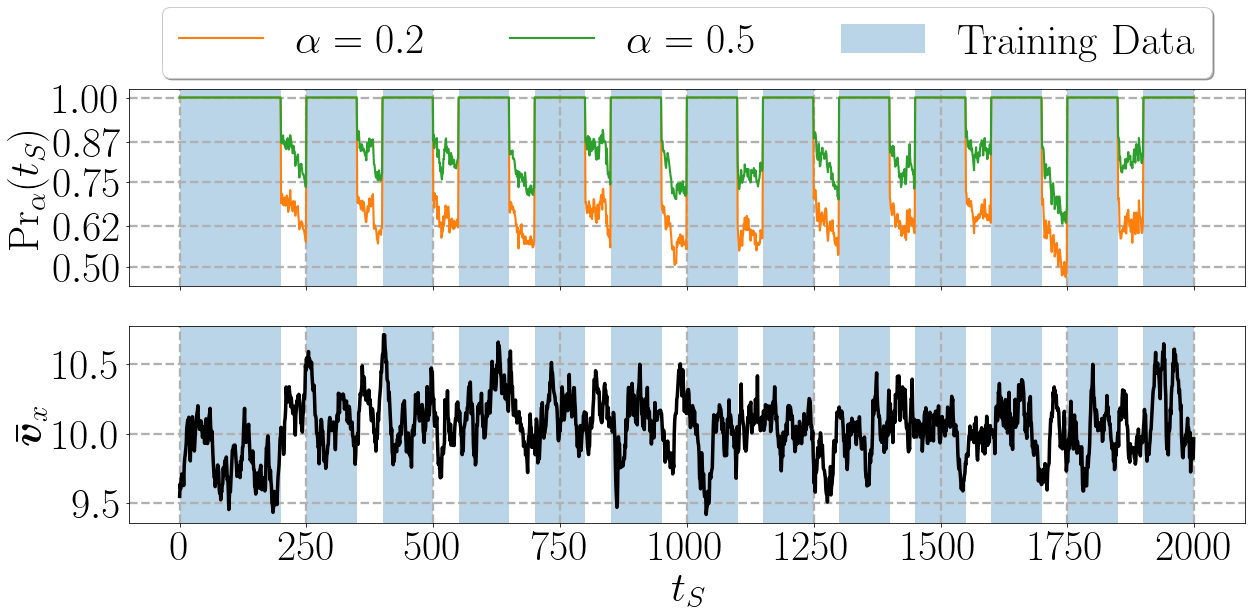}
        \caption{}
	\end{subfigure}
    \hfill
	\begin{subfigure}[b]{0.49\textwidth}
		\centering
		\includegraphics[width=\textwidth]{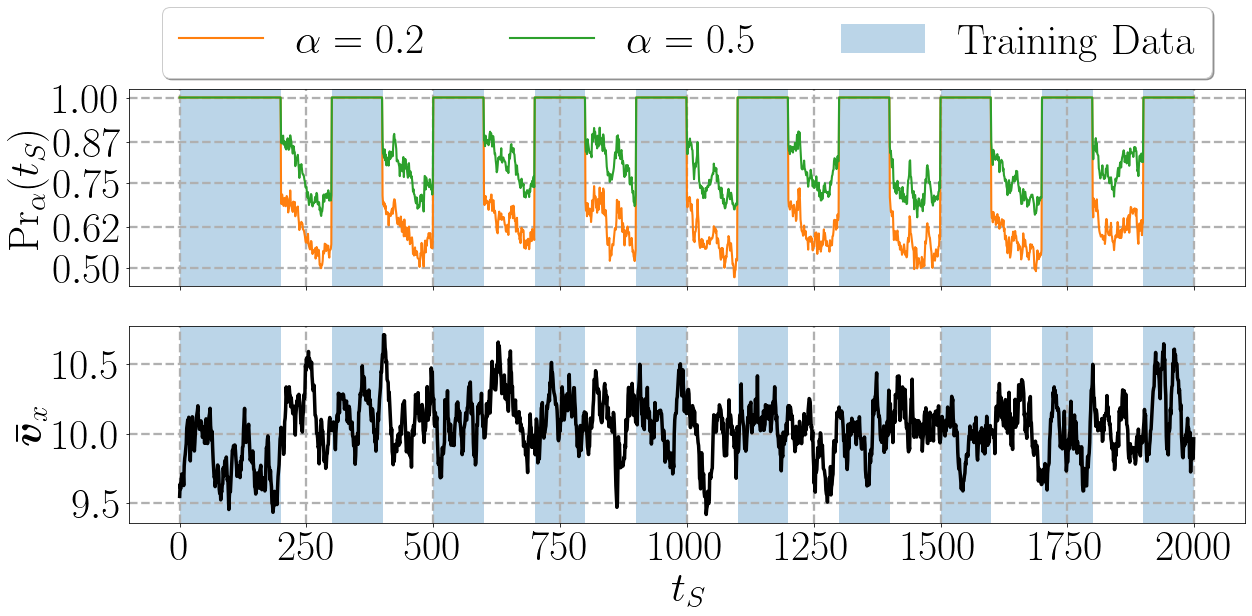}
        \caption{}
	\end{subfigure}
    \hfill
	\begin{subfigure}[b]{0.49\textwidth}
		\centering
		\includegraphics[width=\textwidth]{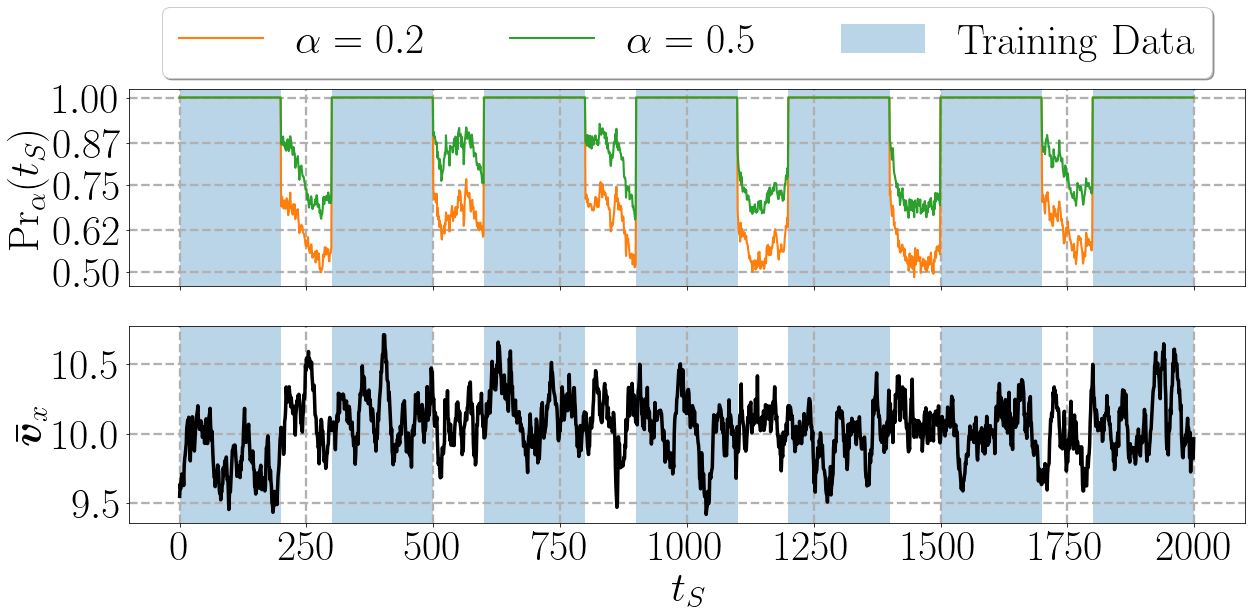}
        \caption{}
	\end{subfigure}
    \hfill
	\begin{subfigure}[b]{0.49\textwidth}
		\centering
		\includegraphics[width=\textwidth]{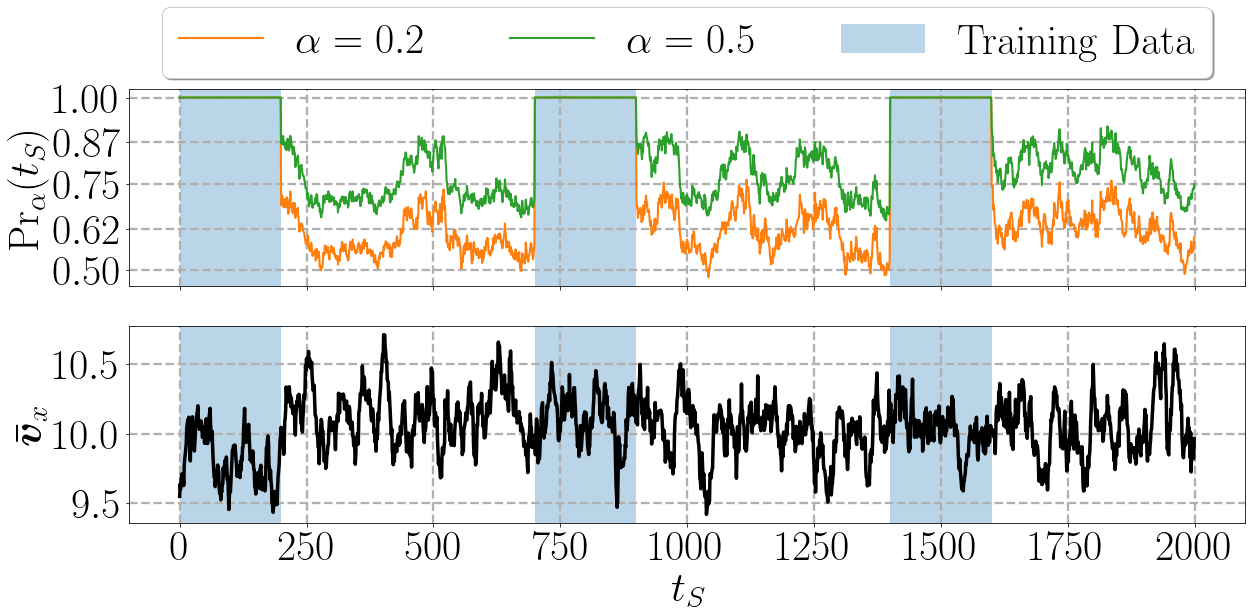}
        \caption{}
	\end{subfigure}
	\caption{The top row of all figures displays the probability of achieving an absolute vorticity prediction error below thresholds of $0.5$ (orange) and $1$ (green) for the four test cases: T1 (a), T2 (b), T3 (c) and T4 (d), in the turbulent flow. The bottom row presents the temporal evolution of the spatially averaged streamwise velocity. These error thresholds are selected based on the typical range of vorticity values identified in the flow, as evident in Fig. \ref{fig: vki_snap}.}
	\label{fig: vki_prob_err}
\end{figure}

Figure \ref{fig: vki_prob_err} shows the probability of achieving an absolute vorticity prediction error less than or equal to $0.2$ and $0.5$, respectively, across the four test cases described in Tab. \ref{tab: cyl_vki_hyperparameters_cases}. This probability is obtained by the PPE metric, defined in eq. \eqref{eq: ppe}, and represented by the probability $\text{Pr}_{\alpha}(t_{S}) \equiv \text{Pr}(-\alpha \leq \boldsymbol{\omega}_{s} - \boldsymbol{\hat{\omega}}_{s} \leq \alpha)$ for $\alpha \in \{0.2, 0.5\}$, where $\boldsymbol{\omega}_{s}$ and $\boldsymbol{\hat{\omega}}_{s}$ denote the $s$-th ground truth and predicted vorticity snapshots, respectively.

Note that, for the turbulent flow, the spatially averaged streamwise velocity ($\boldsymbol{\bar{v}}_{x}$, bottom row) exhibits more abrupt variations in its temporal evolution, reflecting the high complexity and strongly fluctuating nature of this flow.

Note from Fig. \ref{fig: vki_prob_err} that the probability of achieving an absolute prediction error less than or equal to $\alpha = 0.5$ remains above $70\%$ for most of the prediction horizon, dropping only at isolated points. This indicates that the predictions produced by the POD-DL model generally remain close to the reference dynamics and do not exhibit significant divergence. A similar trend is observed for $\alpha = 0.2$, where the probability remains above $50\%$. Given that a relative error corresponding to $\alpha = 0.2$ is considered very small in the context of a turbulent flow such as the one studied here, maintaining a probability above $50\%$ may indicate that the POD-DL model is effectively capturing the overall dynamics of the flow.

\begin{figure}[H]
	\centering
	\begin{subfigure}[b]{0.35\textwidth}
		\centering
		\includegraphics[width=\textwidth]{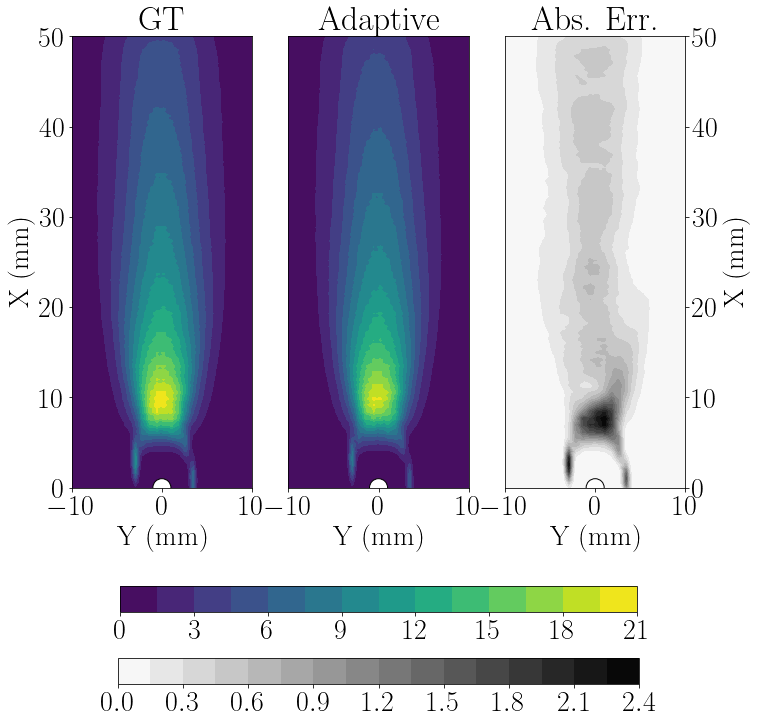}
        \caption{}
	\end{subfigure}
    \hspace{2mm}
	\begin{subfigure}[b]{0.35\textwidth}
		\centering
		\includegraphics[width=\textwidth]{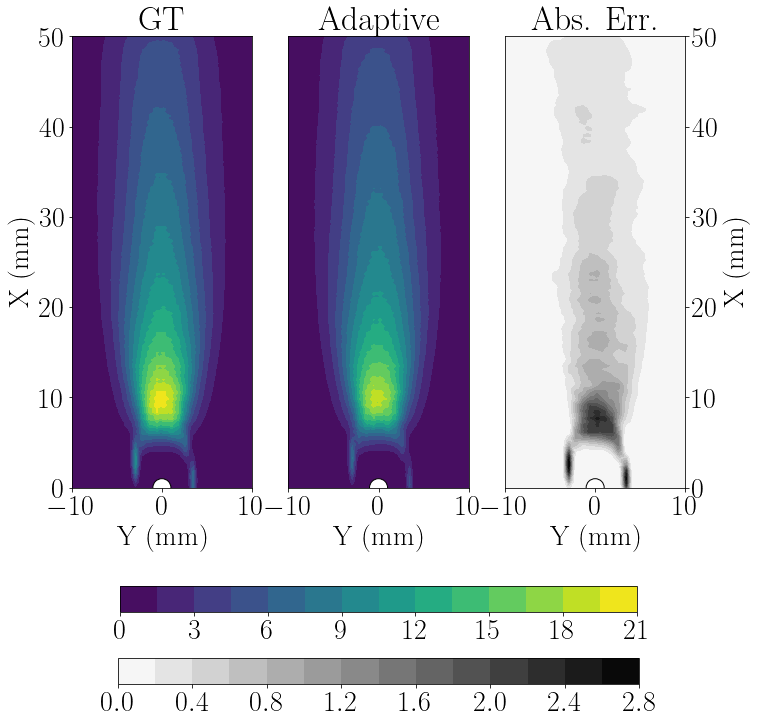}
        \caption{}
	\end{subfigure}
    \hspace{2mm}
	\begin{subfigure}[b]{0.35\textwidth}
		\centering
		\includegraphics[width=\textwidth]{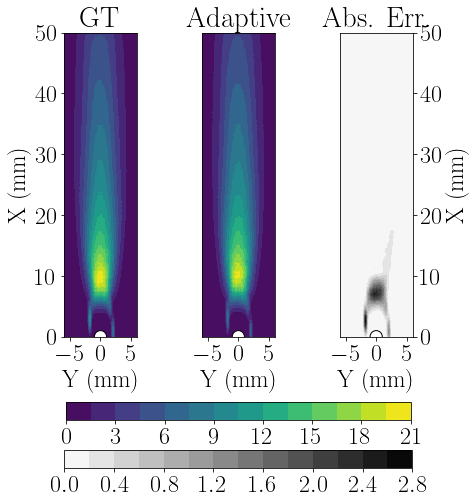}
        \caption{}
	\end{subfigure}
    \hspace{2mm}
	\begin{subfigure}[b]{0.35\textwidth}
		\centering
		\includegraphics[width=\textwidth]{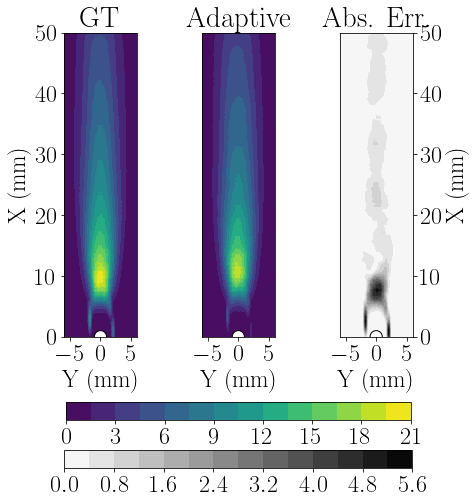}
        \caption{}
	\end{subfigure}
	\caption{Turbulent kinetic energy (TKE) from the ground truth (left) and adaptive framework (middle) for test cases T1 (a), T2 (b), T3 (c), and T4 (d) in the turbulent flow, with the right panels showing the absolute prediction error.}
	\label{fig: vki_tke}
\end{figure}

This observation is further supported by comparing the TKE, computed as defined in eq. \eqref{eq: tke}, between the reference dynamics and the dynamics predicted by the adaptive framework. This comparison is presented in Fig. \ref{fig: vki_tke}, where the right column shows that the absolute error remains small relative to the overall range of energy values. The largest discrepancy is observed in test case T4, which corresponds to the scenario with the highest number of predicted time steps.

\begin{figure}[H]
	\centering
	\begin{subfigure}[b]{0.49\textwidth}
		\centering
		\includegraphics[width=\textwidth]{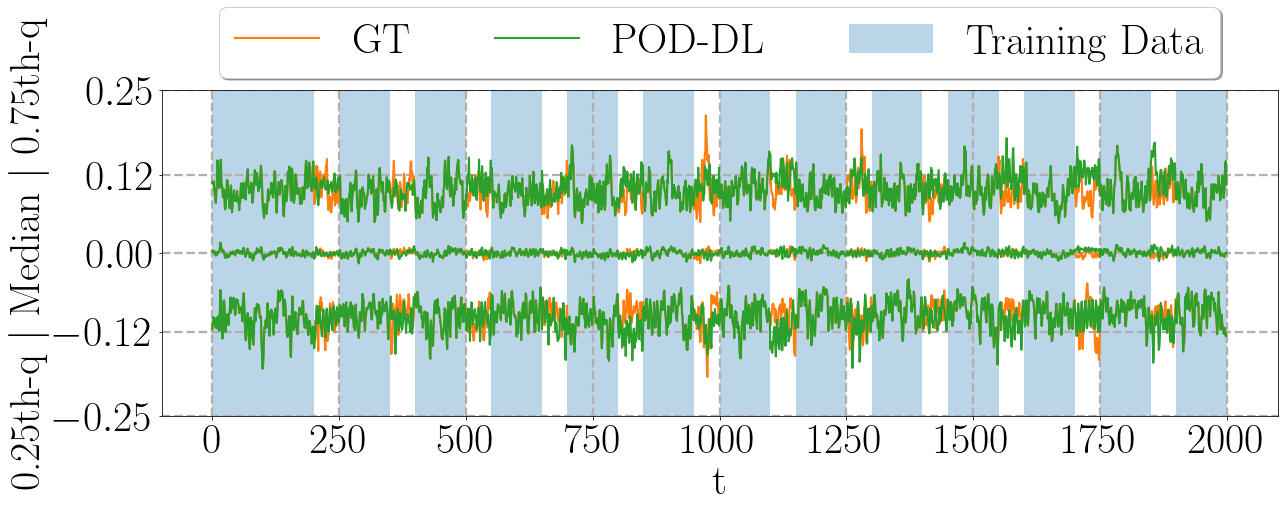}
        \caption{}
	\end{subfigure}
    \hfill
	\begin{subfigure}[b]{0.49\textwidth}
		\centering
		\includegraphics[width=\textwidth]{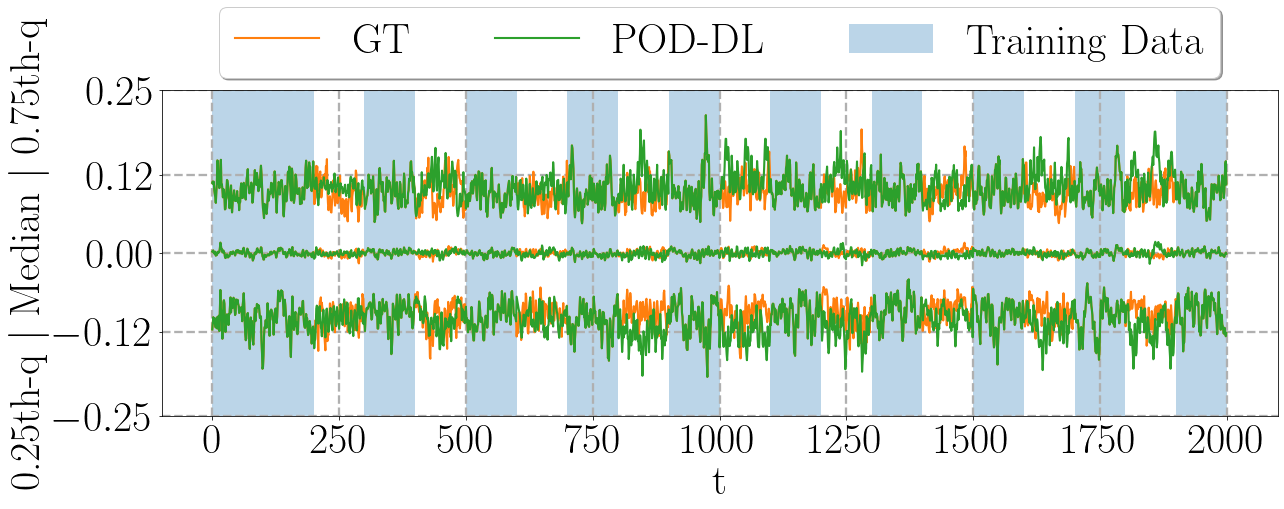}
        \caption{}
	\end{subfigure}
    \hfill
	\begin{subfigure}[b]{0.49\textwidth}
		\centering
		\includegraphics[width=\textwidth]{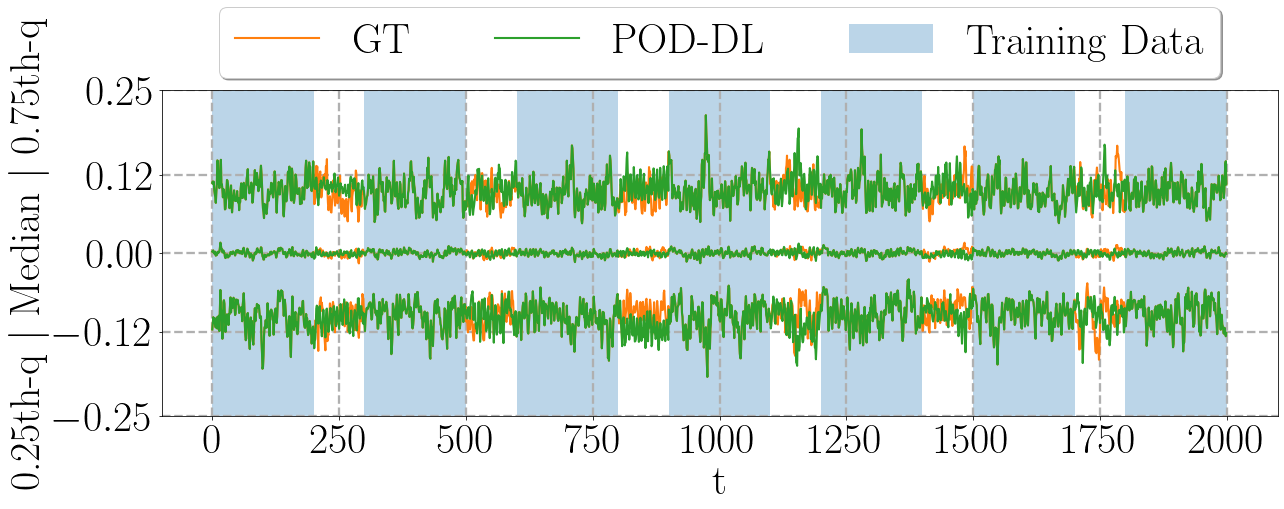}
        \caption{}
	\end{subfigure}
    \hfill
	\begin{subfigure}[b]{0.49\textwidth}
		\centering
		\includegraphics[width=\textwidth]{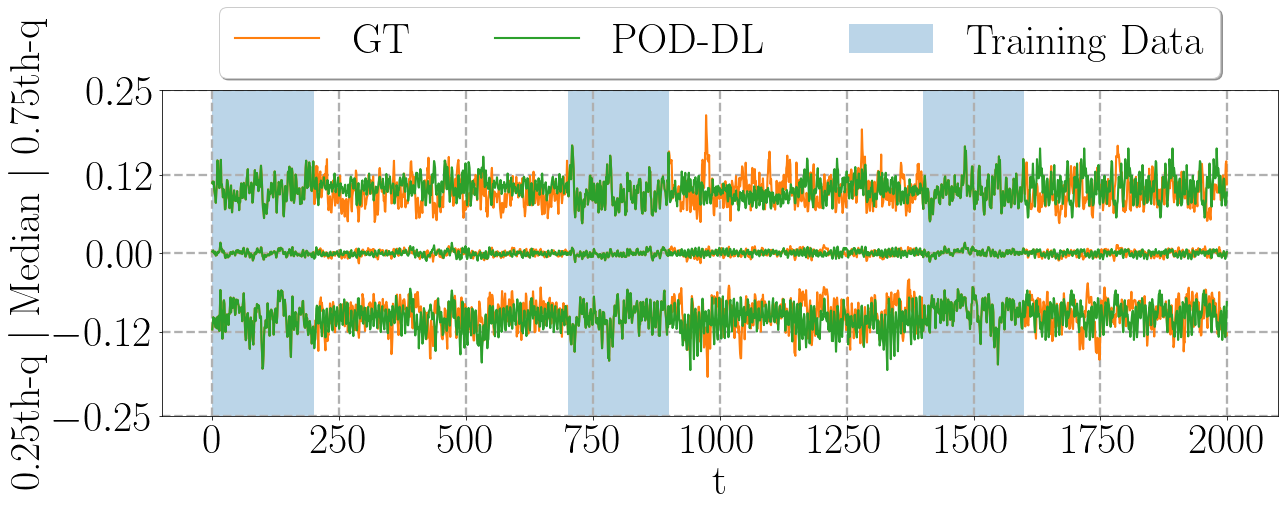}
        \caption{}
	\end{subfigure}
	\caption{Temporal evolution of the 25th (bottom), 50th (middle), and 75th (top) vorticity percentiles from ground truth (blue) and POD-DL predictions (orange) for test cases T1 (a), T2 (b), T3 (c), and T4 (d) in the turbulent flow scenario. Shaded regions indicate training snapshots used in both the initial and retraining phases.}
	\label{fig: vki_quantiles}
\end{figure}

The observation that predictions remain consistent with the underlying physics is also supported by visual inspection of snapshots comparing the predicted and actual flow dynamics, as illustrated in Fig. \ref{fig: vki_snap}. Additionally, Fig. \ref{fig: vki_quantiles} presents a comparison of the 25th, 50th, and 75th percentiles computed for each snapshot. This figure shows that the predicted percentiles closely follow the trends of the ground truth dynamics, which is particularly important in highly complex flows where achieving exact pointwise predictions is impractical.

All test cases studied for the turbulent flow reveal that, even in highly complex dynamics, the adaptive framework successfully adapts to variations in the dynamics. This is accomplished by leveraging new data from the ground truth dynamics and employing transfer learning, which utilizes the weights from previous training iterations to enhance the subsequent retraining.

The computation time required to train the POD-DL model with $13$ POD coefficients and a training set of $200$ samples is approximately $1.76$ minutes on a CPU Intel(R) Core(TM) i7-10700 CPU @ 2.90GHz with $16$ cores and $64$ Gb of RAM. Similar to the laminar flow case, the computation time for predictions remains nearly negligible, averaging around $2$ seconds with the same hardware. In this context, the theoretical computational saving achieved, measured by TCS in eq. \eqref{eq: tcs}, varies across test cases: for test cases T1 and T3, it averages $30\%$; for test case T2, it is around $45\%$; and for test case T4, it is about $70\%$.

\subsection{Isothermal subsonic jet}\label{sec: results_jet_les}

In contrast to the other two cases, where the datasets defines a flow past a circular cylinder at different Reynolds numbers, respectively. In this section the dataset employed to test the adaptive framework is derived from a numerical simulation of an isothermal subsonic jet \cite{guillaume_etal_2018_Jet_LES} issuing from a round nozzle with an exit diameter of $D = 5$ millimeter (mm).

The simulation was conducted using the CharLES large eddy simulation (LES) flow solver, which solves the spatially filtered compressible Navier-Stokes equations on unstructured grids through a combination of a finite-volume approach and third-order Runge-Kutta time integration. The simulation involved approximately 16 million control volumes and covered $2000$ acoustic time units ($t c_{\infty} / D$) at Re $\approx 10^{6}$. In this case $c_{\infty}$ represents the free-stream constant coefficient for the Vreman subgrid model \cite{vreman_2004}.

Specifically, this dataset $\boldsymbol{\mathring{D}}$ consists of $2000$ snapshots of the jet's streamwise velocity, $\boldsymbol{v} = \boldsymbol{v}_{x}$, spanning the temporal domain $\mathcal{P}_{t}$, such that $\boldsymbol{\mathring{D}} \in \mathbb{R}^{1 \times 39 \times 175 \times 1 \times 2000}$. The data was collected on a structured cylindrical output grid designed to closely match the resolution of the LES. This grid covers a streamwise length of $30D$ and a radial extent of $6D$, providing a detailed representation of the jet's spatial dynamics.

Similar to the turbulent flow past a cylinder, the intricate dynamics of this flow led to the exploration of multiple variations of the hyperparameters within the adaptive framework. These variations are detailed in Tab. \ref{tab: jet_hyperparameters_cases}, where the amount of data required for the initial training was again selected to be relatively small compared to the total number of snapshots. However, due to the relative simplicity of the dynamics compared to the turbulent flow, this initial number was reduced to $S_{0} = 100$.

\begin{table}[h]
    \centering
    \caption{Hyperparameters used to test the adaptive framework on the isothermal subsonic jet.\label{tab: jet_hyperparameters_cases}}
    
    \begin{tabular}{lccccc}
        \hline\hline
        \textbf{Test} & \textbf{$P$} & \textbf{$S_{0}$} & \textbf{$S_{1}$} & \textbf{$E$}\\
        \hline
        $T1$ & $100$ & $100$ & $50$ & $1500$ \\
        $T2$ & $200$ & $100$ & $100$ & $1500$ \\
        $T3$ & $600$ & $100$ & $100$ & $1500$ \\
        $T4$ & $2000$ & $100$ & $100$ & $1500$ \\
        \hline\hline
    \end{tabular}

\end{table}

\begin{figure}[H]
    \centering
    \includegraphics[width=0.4\textwidth]{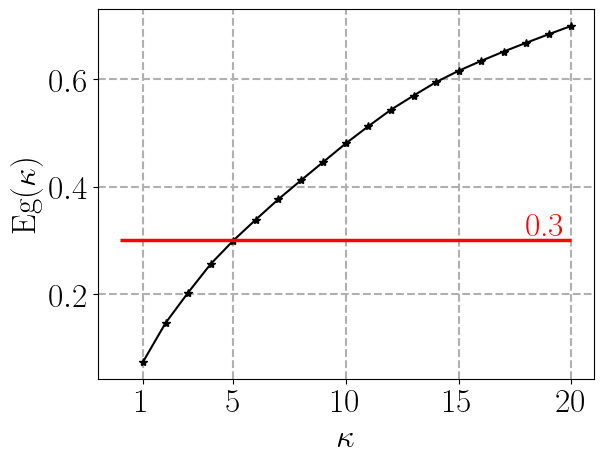}
    \caption{Percentage of energy contained, Eg$(k)$, when the truncation is done at the $n$, up to $20$, most energetic modes in the isothermal jet.}
    \label{fig: jet_sing_val}
\end{figure}

Similar to the previous scenarios, the POD is first applied to the initial $S_{0} = 100$ snapshots to compute the corresponding POD coefficients. The number of POD modes retained is determined based on the amount of energy to be preserved within the dynamics, using the cumulative energy function \eqref{eq: cummEnerg}. Following similar studies \cite{pod_40_energy_causality}, for this type of dynamics it is enough to retain approximately $30\%$ of the total energy to capture the dominant dynamics of interest. As depicted in Fig. \ref{fig: jet_sing_val}, for the first $S_{0} = 100$ snapshots of the isothermal jet this energy threshold is met by preserving the first $5$ modes.


This truncation to $5$ modes is consistently applied throughout all training iterations within the adaptive framework to maintain architectural consistency with the POD-DL model. Unlike the other cases, this dataset contains only the streamwise velocity component; therefore, the model is both trained and evaluated directly on the velocity field.

\begin{figure}[H]
	\centering
	\begin{subfigure}[b]{0.28\textwidth}
		\centering
		\includegraphics[width=\textwidth]{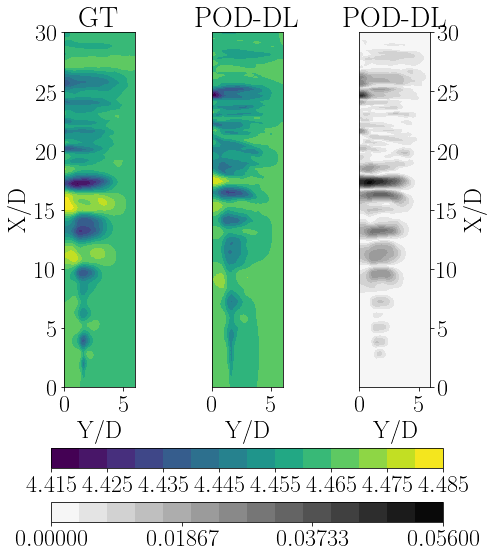}
        \caption{$t_{S} = 1490$}
	\end{subfigure}
    \hfill
	\begin{subfigure}[b]{0.28\textwidth}
		\centering
		\includegraphics[width=\textwidth]{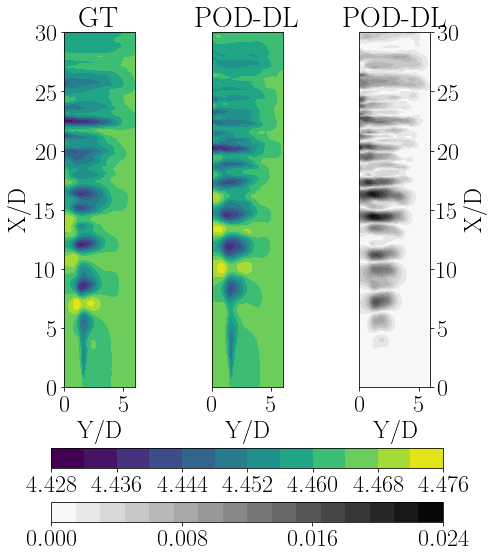}
        \caption{$t_{S} = 750$}
	\end{subfigure}
    \hfill
	\begin{subfigure}[b]{0.28\textwidth}
		\centering
		\includegraphics[width=\textwidth]{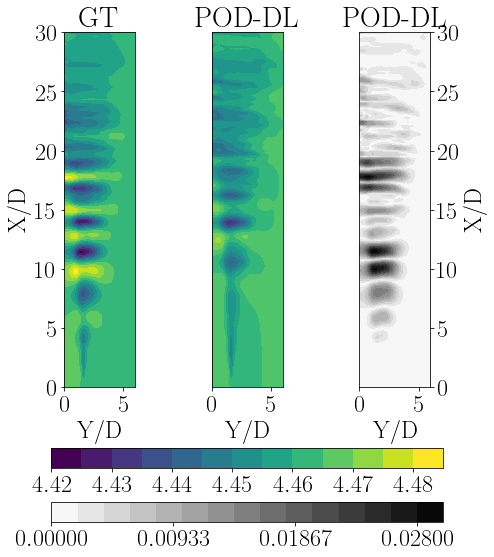}
        \caption{$t_{S} = 1750$}
	\end{subfigure}
	\caption{Snapshots comparing the ground truth (GT) vorticity, of the laminar flow past a cylinder, with the one predicted from POD-DL, at some representative time steps for test cases T1 (a), T2 (b) and T4 (c), respectively. Here $t_{S}$ indicates the snapshot index rather than the actual time $t$.}
	\label{fig: jet_snap}
\end{figure}

As in the previous two flow scenarios, Fig. \ref{fig: jet_snap} presents a visual comparison between the predicted and ground truth velocity fields at selected time steps, for the isothermal jet. It is evident from this figure that the range of velocity values in the isothermal jet is significantly narrower than in the other cases, lying approximately within the interval [$4.405$, $4.495$]. Consequently, an absolute prediction error of $0.02$ corresponds to a relative error range of $[0.02 / 4.495, 0.02 / 4.405] \times 100\% = [0.44\%, 0.45\%]$. Similarly, an absolute error of $0.01$ results in a relative error range of $[0.22\%, 0.23\%]$. These error magnitudes are notably small, implying that predictions with high probabilities of achieving absolute errors below these thresholds are highly reliable.

\begin{figure}[H]
	\centering
	\begin{subfigure}[b]{0.49\textwidth}
		\centering
		\includegraphics[width=\textwidth]{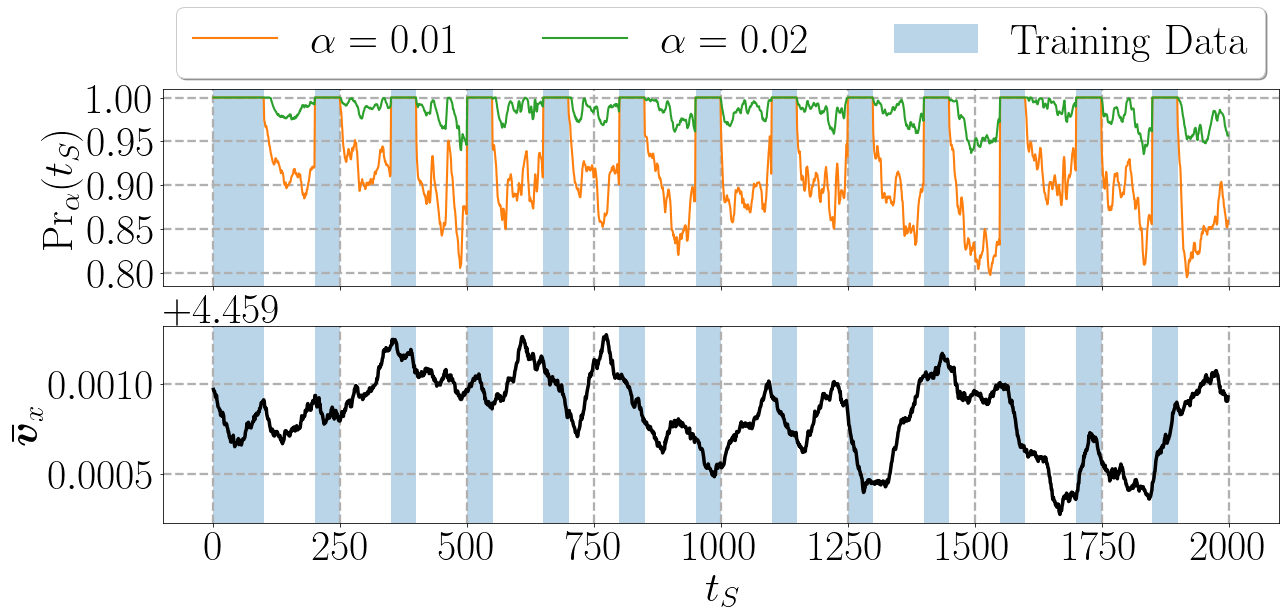}
        \caption{}
	\end{subfigure}
    \hfill
	\begin{subfigure}[b]{0.49\textwidth}
		\centering
		\includegraphics[width=\textwidth]{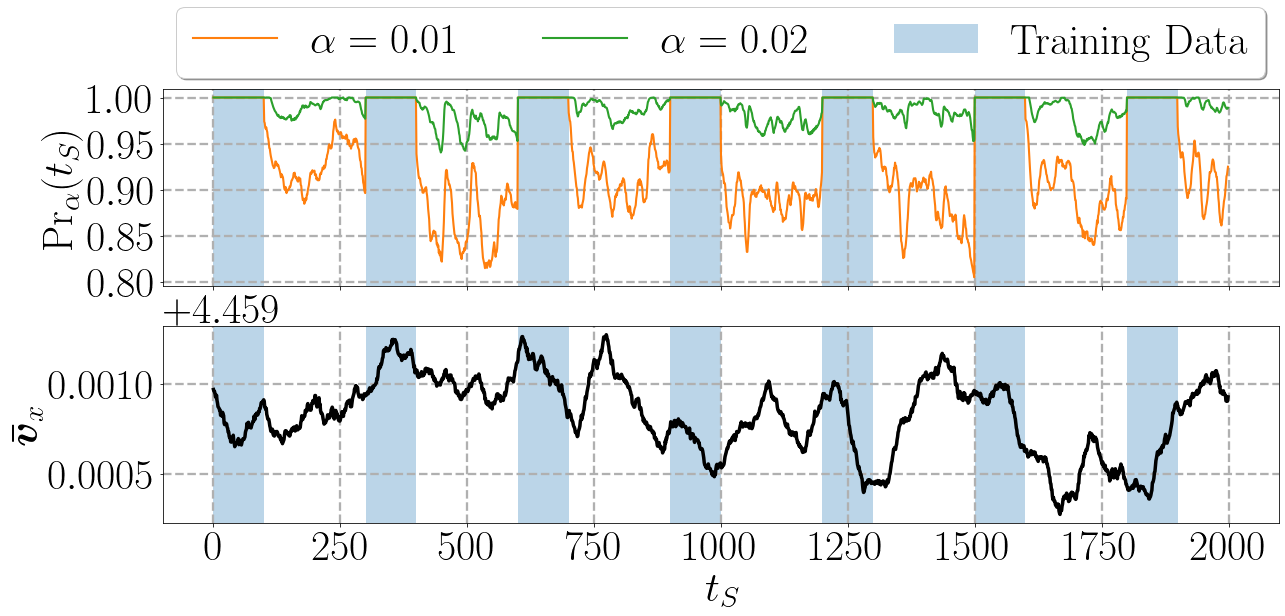}
        \caption{}
	\end{subfigure}
    \hfill
	\begin{subfigure}[b]{0.49\textwidth}
		\centering
		\includegraphics[width=\textwidth]{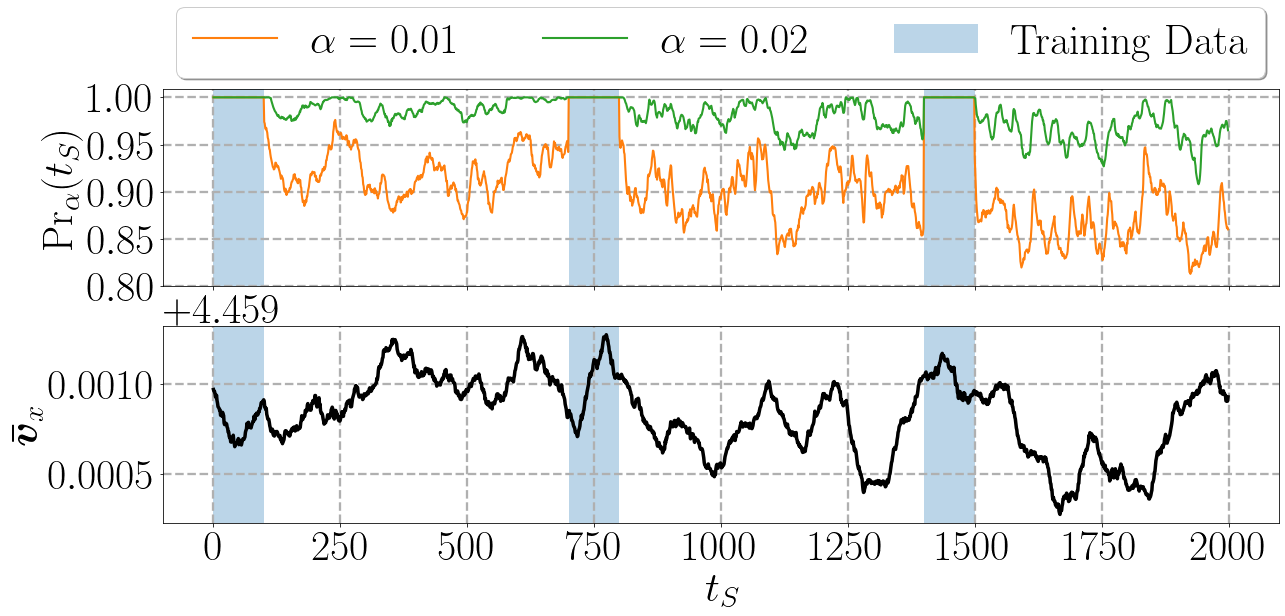}
        \caption{}
	\end{subfigure}
    \hfill
	\begin{subfigure}[b]{0.49\textwidth}
		\centering
		\includegraphics[width=\textwidth]{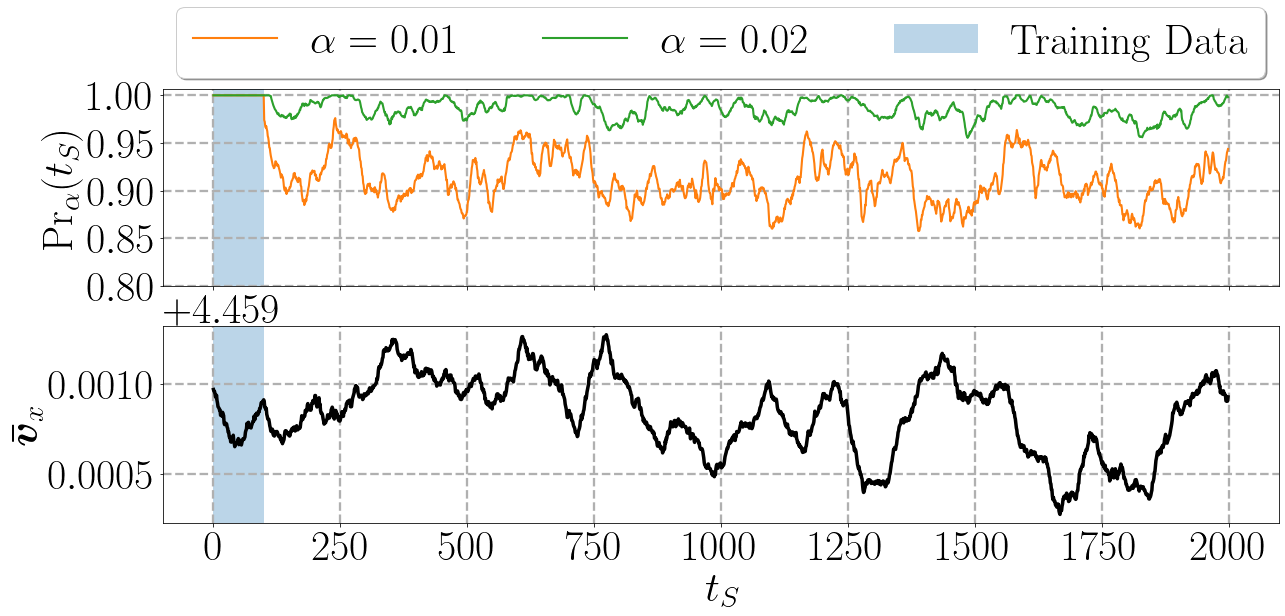}
        \caption{}
	\end{subfigure}
	\caption{The top row of all figures displays the probability of achieving an absolute vorticity prediction error below thresholds of $0.01$ (orange) and $0.02$ (green) for the four test cases: T1 (a), T2 (b), T3 (c) and T4 (d), in the isothermal jet. The bottom row presents the temporal evolution of the spatially averaged streamwise velocity. These error thresholds are selected based on the typical range of vorticity values identified in the flow, as evident in Fig. \ref{fig: jet_snap}.}
	\label{fig: jet_prob_err}
\end{figure}

These probabilities are represented in Fig. \ref{fig: jet_prob_err}, which shows the probability of achieving an absolute velocity prediction error less than or equal to $0.01$ and $0.02$, respectively, across the four test cases described in Tab. \ref{tab: jet_hyperparameters_cases}. This probability is obtained by the PPE metric, defined in equation \ref{eq: ppe}, and represented by the probability $\text{Pr}_{\alpha}(t_{S}) \equiv \text{Pr}(-\alpha \leq \boldsymbol{v}_{s} - \boldsymbol{\hat{v}}_{s} \leq \alpha)$ for $\alpha \in \{0.01, 0.02\}$, where $\boldsymbol{v}_{s}$ and $\boldsymbol{\hat{v}}_{s}$ denote the $s$-th ground truth and predicted velocity snapshots, respectively.

As shown in Fig. \ref{fig: jet_snap}, the range of velocity values in the isothermal jet is noticeably shorter than in the other cases, lying approximately within the interval [$4.405$, $4.495$]. Therefore, an absolute prediction error of $0.02$ in the velocity is negligible for this flow. As such, an model exhibiting a high probability ($\text{Pr}_{\alpha}(t_{S}) \simeq 1$) of producing errors below this threshold indicates highly accurate model. As illustrated in Fig. \ref{fig: jet_prob_err}, this condition is typically satisfied in all predictions performed by the POD-DL model, being the exception at some isolated points.

The same figure also shows the probability of achieving an absolute prediction error less than or equal to $0.01$, which is considered negligible. Notably, this probability remains above $80\%$ throughout the entire prediction horizon. This suggests that the predictions made by the POD-DL model stay closely aligned with the reference dynamics and do not exhibit significant divergence.

We believe these results are largely attributable to the fact that the flow dynamics in the isothermal jet are statistically steady, which means that it has already reached a statistically converged state, with no significant distributional changes occurring over time. This is evident in the evolution of the spatially averaged streamwise velocity, displayed in the bottom row of all figures in Fig. \ref{fig: jet_prob_err}, where the variation is nearly negligible. Such stability enables the POD-DL model to effectively capture the overall trend of the dynamics with minimal training data.

\begin{figure}[H]
	\centering
	\begin{subfigure}[b]{0.49\textwidth}
		\centering
		\includegraphics[width=\textwidth]{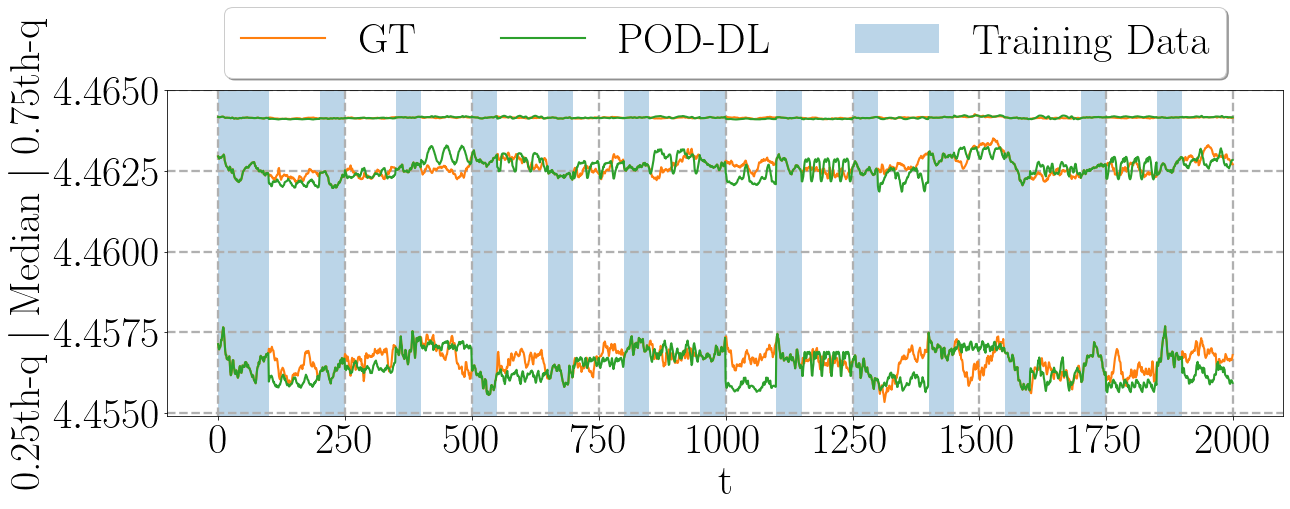}
        \caption{}
	\end{subfigure}
    \hfill
	\begin{subfigure}[b]{0.49\textwidth}
		\centering
		\includegraphics[width=\textwidth]{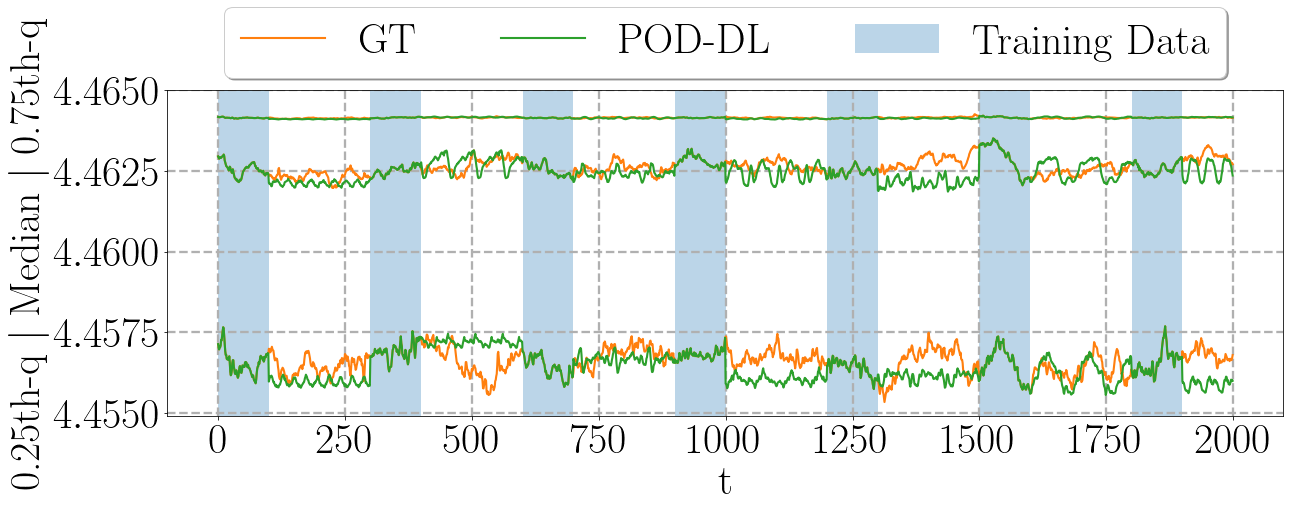}
        \caption{}
	\end{subfigure}
    \hfill
	\begin{subfigure}[b]{0.49\textwidth}
		\centering
		\includegraphics[width=\textwidth]{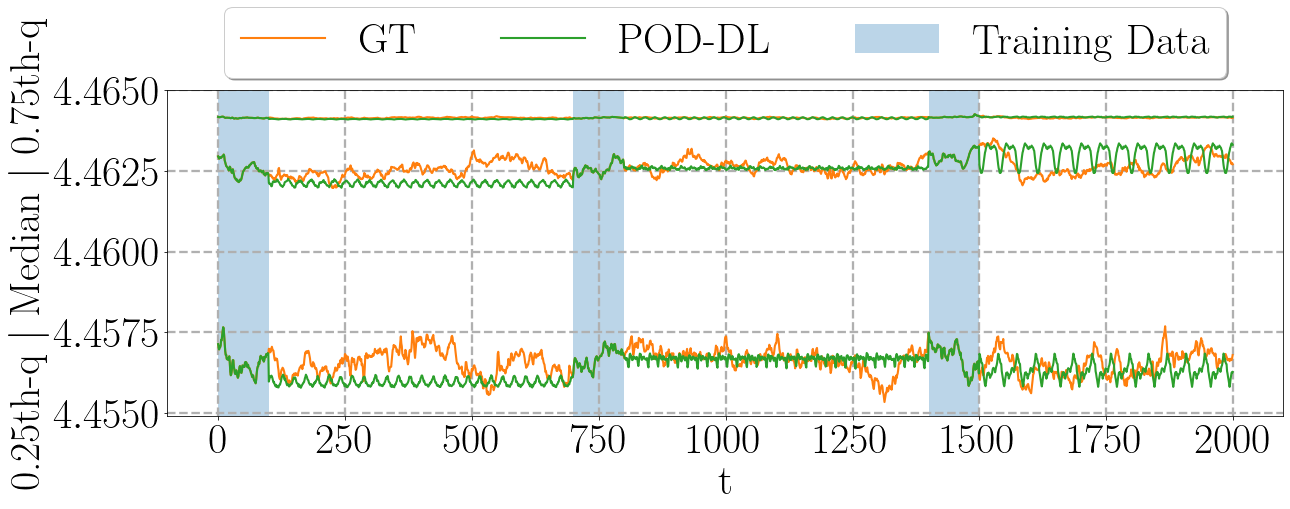}
        \caption{}
	\end{subfigure}
    \hfill
	\begin{subfigure}[b]{0.49\textwidth}
		\centering
		\includegraphics[width=\textwidth]{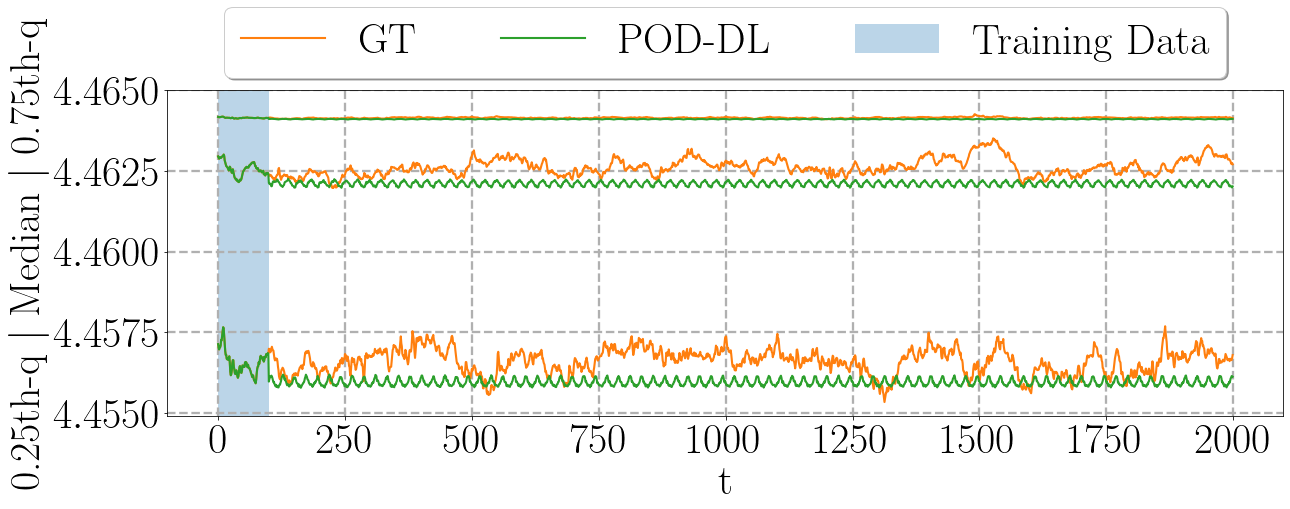}
        \caption{}
	\end{subfigure}
	\caption{Temporal evolution of the 25th (bottom), 50th (middle), and 75th (top) vorticity percentiles from ground truth (blue) and POD-DL predictions (orange) for test cases 1 (a), 2 (b), 3 (c), and 4 (d) in the turbulent flow scenario. Shaded regions indicate training snapshots used in both the initial and retraining phases.}
	\label{fig: jet_quantiles}
\end{figure}

This is further illustrated in Fig. \ref{fig: jet_quantiles}, which compares the $25$th, $50$th, and $75$th percentiles obtained from the ground truth dynamics and the adaptive framework. The variation in these percentile values is minimal, and while the POD-DL model generally follows the overall trend, it struggles to capture these subtle fluctuations. However, it is noteworthy that in test cases T1 and T2, where more training data are incorporated, the model's predictions more closely approximate these variations. This suggests that the recurrent inclusion of new training data enhances the POD-DL model's ability to adapt to even small-scale fluctuations in the flow.

The computation time required to train the POD-DL model with $5$ POD coefficients and a training set of $100$ samples is approximately $1.18$ minutes on a CPU Intel(R) Core(TM) i7-10700 CPU @ 2.90GHz with $16$ cores and $64$ Gb of RAM. Similar to the other flow cases, the computation time for predictions remains nearly negligible, averaging around $2$ seconds with the same hardware. In this context, the theoretical computational savings, as defined in eq. \eqref{eq: tcs}, vary across the different test cases: for T1 and T2 the average saving is approximately $65\%$, for T3 it reaches around $85\%$ and for T4 it is about $95\%$. These results underscore the effectiveness of the adaptive framework in significantly reducing the computational cost associated with traditional numerical solvers, while preserving a high level of predictive accuracy.

Next section provides a general discussion about the results obtained from the adaptive framework and discuss future lines of work.

\section{Discussion}\label{sec: discussion}

The adaptive framework proposed in this study has been evaluated on three datasets, each representing different flow dynamics at varying Reynolds numbers. The results, presented in Sec. \ref{sec: results}, indicate that the highest accuracy was achieved for the datasets corresponding to the laminar flow past a circular cylinder (Sec. \ref{sec: results_laminar_cyl}) and the isothermal subsonic jet (Sec. \ref{sec: results_jet_les}). The superior performance of the POD-DL model, in forecasting the flow dynamics of the isothermal jet, can be attributed to the statistically steady regime attained by the flow, even under turbulent conditions, which ensures that the data distribution remains consistent over time. This is visually evident in the bottom row of all figures in Fig. \ref{fig: jet_prob_err}, which depicts the temporal evolution of the spatial average of the streamwise velocity. Notably, the variation in this quantity is almost negligible.

Since the data distribution remains unchanged, the temporal correlations within the sequence also remain consistent. This implies that, with a sufficiently large training dataset to capture these temporal correlations, the model can accurately extrapolate the evolution of the POD coefficients and, consequently, the flow dynamics.

However, as observed in the laminar flow past a cylinder, when the data distribution begins to change, the predictions start to diverge since the model has not been informed of this shift. This effect is clearly illustrated in Fig. \ref{fig: cil_prob_err}, for test cases T1 (a) and T3 (c), where predictions fail to capture a variation in the dynamics, leading to divergence from the reference solution.

The adaptive framework has been introduced as a solution to address this issue of distribution shifts in the dynamics. In this approach, the POD-DL model is retrained with newly acquired data, allowing it to account for unseen changes in the data distribution over time. This effect is demonstrated in Fig. \ref{fig: cil_prob_err} for test case T2 (b), where a successive retraining of the POD-DL model, informing about the variation in the dynamics, results in long term stable and accurate predictions.

Special attention should be given to Fig. \ref{fig: cil_prob_err} (a), where it can be observed that, even after the second retraining, the predictions begin to diverge. This occurs because the newly incorporated training data does not contain sufficient information about the impending change in the dynamics, limiting the model's ability to accurately forecast the evolution of the flow. However, once the model undergoes a third retraining with additional data corresponding to a more stable phase of the dynamics, the predictions regain consistency. Despite this intermediate divergence, the adaptive framework ultimately enables the POD-DL model to produce accurate predictions of the flow dynamics.

These cases further emphasize the importance of an adaptive framework that allows the model to remain flexible to changes in the data distribution, a crucial feature for addressing complex problems such as in fluid dynamics.

For instance, the most complex case analyzed in this work, which corresponds to the three-dimensional wake of a circular cylinder at turbulent regime (Sec. \ref{sec: results_turbulent_cyl}), exhibits frequent and significant variations in the evolution of the spatial mean velocity (bottom row in all figures of Fig. \ref{fig: vki_prob_err}), which are associated with the presence of small-scale structures in the dynamics that display nearly chaotic behavior. These rapid fluctuations considerably reduce the time horizon over which the POD-DL model can produce accurate predictions without divergence. This limitation is evident across all studied configurations of the adaptive framework. Nevertheless, incorporating new data for retraining allows the model to regain predictive accuracy and restore convergence.

As illustrated in the same figure, the POD-DL model is capable of delivering reliable predictions during the initial time steps following each retraining phase. Beyond this threshold, the predictions begin to diverge rapidly, highlighting the challenge of capturing the distributions shifts that are constantly happening in the dynamics.

However, even in this highly complex case, the adaptive framework demonstrates its ability to mitigate errors through retraining, enabling the POD-DL model to adapt to previously unseen changes in the dynamics while remaining consistent with the underlying physics of the problem. This is evidenced in Fig. \ref{fig: vki_tke}, which compares the turbulent kinetic energy of the ground truth dynamics with that obtained from the adaptive framework.

One of the primary limitations of the adaptive framework lies in the deep learning model used to predict the temporal evolution of the POD coefficients. As observed in complex scenarios such as the turbulent flow past a circular cylinder, the model may exhibit a limited prediction window and high sensitivity to shifts in the data distribution, which constrains its extrapolation capability. The DL model employed in this work, however, is computationally efficient and, due to its relative simplicity, offers improved interpretability. Nevertheless, alternative deep learning architectures could be explored, focusing on models that better capture intricate temporal correlations or demonstrate greater robustness to changes in data distribution. Potential options include extended long short-term memory (xLSTM) networks \cite{xlstm}, an extension of the standard LSTM used in this work, which aims to improve long-range dependency modeling. Another promising approach is multi-layer perceptron (MLP) mixer models \cite{temporal_mixer}, a recent research direction that investigates the effectiveness of fully MLP-based architectures for forecasting tasks. Additionally, diffusion models \cite{Lin_etal_2024_diffusion_forecasting_survey} offer an alternative paradigm by approximating the underlying probability distribution of the forecasting process.

In the case of diffusion models, there is a growing interest in the literature in evaluating their ability to capture extreme events within datasets \cite{diffusion_models_extreme_events}. This is particularly relevant to the present study, where the forecasting model should extrapolate unseen events that were not present during training. Such models, which can accurately represent extreme events in statistical distributions, have the potential to enhance prediction accuracy in complex scenarios, such as the turbulent flow past a cylinder. But exploring the performance of these techniques within this adaptive framework remains open for future research. This study was conducted as a proof of concept to demonstrate the feasibility of an adaptive framework for accelerating CFD simulations, with a primary focus on evaluating the potential computational savings it can provide.

Other components have also been left for future research, one such area is the implementation of an uncertainty quantification method capable of dynamically truncating the number of predictions based on their reliability, without requiring direct comparison with ground truth data. Several existing studies propose potential approaches to address this challenge. Examples include regular bootstrap techniques \cite{uq_adaled, adaled_petros}, methods that exploit the structure of POD \cite{rapun_adaptive_pod_modes_2015, beltran_adaptive, uq_abadiaheredia_etal}, conformal prediction frameworks \cite{uq_conformal_prediction}, and approaches based on evaluating the residuals of the governing PDEs \cite{rapun_adaptive_pod_modes_2015}.

Another future research direction is the integration of a numerical solver that can be invoked when additional data is required for retraining the POD-DL model. In this study, this component has been replaced by retrieving data from a precomputed dataset to supply new training samples.

It is important to note that incorporating a numerical solver within the adaptive framework may reduce the computational speedup achieved, as each solver execution begins with the last snapshot predicted by the POD-DL model. This setup inherently requires a certain number of time steps for the solver to reach convergence. However, in practice, this number is typically small, suggesting that the trade-off between accuracy and computational efficiency could still be favorable.

\section{Conclusions}\label{sec: conclusions}
This works presents a proof of concept for an adaptive framework designed to accelerate numerical simulations in computational fluid dynamics (CFD) problems. The framework is characterized by the use of a hybrid forecasting model that integrates proper orthogonal decomposition (POD) with forecasting deep learning (DL) models. By applying POD, the method reduces a high-dimensional dataset to a low-dimensional representation, where each time step is expressed as a low-dimensional vector. This dimensionality reduction significantly decreases the complexity of the DL model required, leading to faster training times, a crucial advantage in an adaptive framework aimed at enhancing the efficiency of numerical simulations.

The effectiveness of the proposed adaptive framework has been evaluated across three different fluid dynamics problems, encompassing both laminar and turbulent flows at varying Reynolds numbers. In all cases, the framework demonstrated its ability to reduce the computational time required for simulating flow dynamics compared to conventional CFD solvers.

Furthermore, the accuracy of the predictions was found to be influenced not only by the complexity of the flow dynamics but primarily by the magnitude and frequency of variations in the data distribution. If these variations are not captured during training, the forecasting model will be unable to account for them, necessitating retraining with updated data. Since such distribution shifts are common in fluid dynamics problems, this study underscores the importance of adaptive frameworks like the one proposed in this work, which can dynamically adjust to evolving flow conditions.

This study has not specifically addressed the development and implementation of an uncertainty quantification method capable of assessing the reliability of the predictive model’s forecasts without requiring ground truth data. Such a method would enable the autonomous adjustment of the number of computed predictions, preventing divergence from the actual dynamics. This remains an important avenue for future research.

Additionally, another promising direction is the exploration of DL-based forecasting models that exhibit greater robustness to shifts in the data distribution. Enhancing the model’s adaptability to such variations would not only improve prediction accuracy but also extend the time horizon over which reliable forecasts can be made.

We believe that an adaptive framework like the one proposed in this work has broad potential applications across both industry and academia, offering a versatile approach for accelerating numerical simulations in CFD and related fields.

\section*{Acknowledgments}
The authors acknowledge the grants TED2021-129774B-C21 and PLEC2022-009235 funded by MCIN/AEI/10.13039/501100011033 and by the European Union “NextGenerationEU”/PRTR and the grant PID2023-147790OB-I00 funded by MCIU/AEI/10.13039/501100011033/FEDER, UE. The MODELAIR and ENCODING projects have received funding from the European Union’s Horizon Europe research and innovation programme under the Marie Sklodowska-Curie grant agreement No. 101072559 and 101072779, respectively. The results of this publication reflect only the author(s) view and do not necessarily reflect those of the European Union. The European Union can not be held responsible for them.

\section*{Data Availability Statement}

The code used in this work is openly available at \url{https://github.com/RAbadiaH/adaptive-cfd-forecasting-hybrid-modal-decomposition-deep-learning}. The dataset corresponding to the laminar flow, originally generated by \cite{LeClainche_2018_cilind}, and used to support the findings of this study, can be accessed at \url{https://drive.google.com/drive/folders/1_MkWVuWWoE3hGKPT0FbCba234KJ06kQo} (file: \textit{Tensor.mat}). Additionally, the dataset for the turbulent flow, obtained by \cite{Mendez_2020_experimental}, is available at \url{https://github.com/mendezVKI/MODULO/tree/master/download_all_data_exercises}.

\appendix

\section{Computation of the probability of prediction error} \label{appnx: pape}

Algorithm \ref{alg: pape} shows the pseudocode required to compute the probability of obtaining an absolute prediction error, between a reference ($gt\_ten$) and predicted ($pred\_ten$) snapshots, lower than or equal to an error threshold ($err\_thresh$).

\begin{algorithm} 
\caption{Compute probabilistic prediction error (PPE)}
\label{alg: pape}
\begin{algorithmic}[1]
\Function{compute\_ppe}{$err\_thresh$, $bin\_width$, $gt\_ten$, $pred\_ten$}
    \State $diff \gets$ absolute value of $(gt\_ten - pred\_ten)$
    \State $max\_diff \gets$ maximum value in $diff$
    \State $bin\_edges \gets$ array from $0.0$ to $max\_diff + bin\_width$ with step $bin\_width$
    \State $(p\_hist, \_) \gets$ histogram of $diff$ using $bin\_edges$, normalized to density
    \State $p \gets p\_hist \times bin\_width$
    \State $index \gets \text{integer part of } \frac{err\_thresh}{bin\_width}$
    \State \Return sum of $p$ from bin $0$ to bin $(index - 1)$
\EndFunction
\end{algorithmic}
\end{algorithm}

 \bibliographystyle{elsarticle-harv} 
 \bibliography{references}

\begin{thebibliography}{56}
\expandafter\ifx\csname natexlab\endcsname\relax\def\natexlab#1{#1}\fi
\providecommand{\url}[1]{\texttt{#1}}
\providecommand{\href}[2]{#2}
\providecommand{\path}[1]{#1}
\providecommand{\DOIprefix}{doi:}
\providecommand{\ArXivprefix}{arXiv:}
\providecommand{\URLprefix}{URL: }
\providecommand{\Pubmedprefix}{pmid:}
\providecommand{\doi}[1]{\href{http://dx.doi.org/#1}{\path{#1}}}
\providecommand{\Pubmed}[1]{\href{pmid:#1}{\path{#1}}}
\providecommand{\bibinfo}[2]{#2}
\ifx\xfnm\relax \def\xfnm[#1]{\unskip,\space#1}\fi
\bibitem[{Abadía-Heredia et~al.(2025)Abadía-Heredia, Corrochano, Lopez-Martin and Le~Clainche}]{2025_abadiaheredia_etal_poddl_ae}
\bibinfo{author}{Abadía-Heredia, R.}, \bibinfo{author}{Corrochano, A.}, \bibinfo{author}{Lopez-Martin, M.}, \bibinfo{author}{Le~Clainche, S.}, \bibinfo{year}{2025}.
\newblock \bibinfo{title}{Generalization capabilities and robustness of hybrid models grounded in physics compared to purely deep learning models}.
\newblock \bibinfo{journal}{Physics of Fluids} \bibinfo{volume}{37}, \bibinfo{pages}{035149}.
\newblock \URLprefix \url{https://doi.org/10.1063/5.0253876}, \DOIprefix\doi{10.1063/5.0253876}.
\bibitem[{Abadía-Heredia et~al.(2024)Abadía-Heredia, Lopez-Martín and {le Clainche}}]{uq_abadiaheredia_etal}
\bibinfo{author}{Abadía-Heredia, R.}, \bibinfo{author}{Lopez-Martín, M.}, \bibinfo{author}{{le Clainche}, S.}, \bibinfo{year}{2024}.
\newblock \bibinfo{title}{Evaluating forecast divergence in fluid dynamics: A metric for autoregressive models in the absence of true solution data}, in: \bibinfo{booktitle}{ECCOMAS 2024}, \bibinfo{publisher}{Scipedia}.
\newblock \URLprefix \url{https://www.scipedia.com/public/Abadia-Heredia_et_al_2024a}.
\bibitem[{Abadía-Heredia et~al.(2022)Abadía-Heredia, López-Martín, Carro, Arribas, Pérez and {Le Clainche}}]{abadiaherediaetal_2022}
\bibinfo{author}{Abadía-Heredia, R.}, \bibinfo{author}{López-Martín, M.}, \bibinfo{author}{Carro, B.}, \bibinfo{author}{Arribas, J.}, \bibinfo{author}{Pérez, J.}, \bibinfo{author}{{Le Clainche}, S.}, \bibinfo{year}{2022}.
\newblock \bibinfo{title}{A predictive hybrid reduced order model based on proper orthogonal decomposition combined with deep learning architectures}.
\newblock \bibinfo{journal}{Expert Systems with Applications} \bibinfo{volume}{187}, \bibinfo{pages}{115910}.
\newblock \URLprefix \url{https://www.sciencedirect.com/science/article/pii/S0957417421012653}, \DOIprefix\doi{https://doi.org/10.1016/j.eswa.2021.115910}.
\bibitem[{Aubry(1991)}]{aubry_1991}
\bibinfo{author}{Aubry, N.}, \bibinfo{year}{1991}.
\newblock \bibinfo{title}{On the hidden beauty of the proper orthogonal decomposition}.
\newblock \bibinfo{journal}{Theoret. Comput. Fluid Dynamics} \bibinfo{volume}{2}, \bibinfo{pages}{339--352}.
\newblock \DOIprefix\doi{10.1007/BF00271473}.
\bibitem[{Auer et~al.(2023)Auer, Gauch, Klotz and Hochreiter}]{uq_conformal_prediction}
\bibinfo{author}{Auer, A.}, \bibinfo{author}{Gauch, M.}, \bibinfo{author}{Klotz, D.}, \bibinfo{author}{Hochreiter, S.}, \bibinfo{year}{2023}.
\newblock \bibinfo{title}{Conformal prediction for time series with modern hopfield networks}.
\newblock \URLprefix \url{https://arxiv.org/abs/2303.12783}, \href{http://arxiv.org/abs/2303.12783}{{\tt arXiv:2303.12783}}.
\bibitem[{Beck et~al.(2024)Beck, Pöppel, Spanring, Auer, Prudnikova, Kopp, Klambauer, Brandstetter and Hochreiter}]{xlstm}
\bibinfo{author}{Beck, M.}, \bibinfo{author}{Pöppel, K.}, \bibinfo{author}{Spanring, M.}, \bibinfo{author}{Auer, A.}, \bibinfo{author}{Prudnikova, O.}, \bibinfo{author}{Kopp, M.}, \bibinfo{author}{Klambauer, G.}, \bibinfo{author}{Brandstetter, J.}, \bibinfo{author}{Hochreiter, S.}, \bibinfo{year}{2024}.
\newblock \bibinfo{title}{xlstm: Extended long short-term memory}.
\newblock \URLprefix \url{https://arxiv.org/abs/2405.04517}, \href{http://arxiv.org/abs/2405.04517}{{\tt arXiv:2405.04517}}.
\bibitem[{Beltr{\'a}n et~al.(2020)Beltr{\'a}n, Clainche and Vega}]{2019_beltran_etal_soco}
\bibinfo{author}{Beltr{\'a}n, V.}, \bibinfo{author}{Clainche, S.L.}, \bibinfo{author}{Vega, J.M.}, \bibinfo{year}{2020}.
\newblock \bibinfo{title}{A data-driven rom based on hodmd}, in: \bibinfo{editor}{Mart{\'i}nez~{\'A}lvarez, F.}, \bibinfo{editor}{Troncoso~Lora, A.}, \bibinfo{editor}{S{\'a}ez~Mu{\~{n}}oz, J.A.}, \bibinfo{editor}{Quinti{\'a}n, H.}, \bibinfo{editor}{Corchado, E.} (Eds.), \bibinfo{booktitle}{14th International Conference on Soft Computing Models in Industrial and Environmental Applications (SOCO 2019)}, \bibinfo{publisher}{Springer International Publishing}, \bibinfo{address}{Cham}. pp. \bibinfo{pages}{567--576}.
\bibitem[{Beltrán et~al.(2022)Beltrán, Le~Clainche and Vega}]{beltran_adaptive}
\bibinfo{author}{Beltrán, V.}, \bibinfo{author}{Le~Clainche, S.}, \bibinfo{author}{Vega, J.M.}, \bibinfo{year}{2022}.
\newblock \bibinfo{title}{An adaptive data-driven reduced order model based on higher order dynamic mode decomposition}.
\newblock \bibinfo{journal}{Journal of Scientific Computing} \bibinfo{volume}{92}.
\newblock \URLprefix \url{http://dx.doi.org/10.1007/s10915-022-01855-2}, \DOIprefix\doi{10.1007/s10915-022-01855-2}.
\bibitem[{Besabe et~al.(2025)Besabe, Girfoglio, Quaini and Rozza}]{pod_dl_Besabe_etal_2025}
\bibinfo{author}{Besabe, L.}, \bibinfo{author}{Girfoglio, M.}, \bibinfo{author}{Quaini, A.}, \bibinfo{author}{Rozza, G.}, \bibinfo{year}{2025}.
\newblock \bibinfo{title}{Data-driven reduced order modeling of a two-layer quasi-geostrophic ocean model}.
\newblock \bibinfo{journal}{Results in Engineering} \bibinfo{volume}{25}, \bibinfo{pages}{103691}.
\newblock \URLprefix \url{https://www.sciencedirect.com/science/article/pii/S2590123024019340}, \DOIprefix\doi{https://doi.org/10.1016/j.rineng.2024.103691}.
\bibitem[{Brès et~al.(2018)Brès, Jordan, Jaunet, Le~Rallic, Cavalieri, Towne, Lele, Colonius and Schmidt}]{guillaume_etal_2018_Jet_LES}
\bibinfo{author}{Brès, G.A.}, \bibinfo{author}{Jordan, P.}, \bibinfo{author}{Jaunet, V.}, \bibinfo{author}{Le~Rallic, M.}, \bibinfo{author}{Cavalieri, A.V.G.}, \bibinfo{author}{Towne, A.}, \bibinfo{author}{Lele, S.K.}, \bibinfo{author}{Colonius, T.}, \bibinfo{author}{Schmidt, O.T.}, \bibinfo{year}{2018}.
\newblock \bibinfo{title}{Importance of the nozzle-exit boundary-layer state in subsonic turbulent jets}.
\newblock \bibinfo{journal}{Journal of Fluid Mechanics} \bibinfo{volume}{851}, \bibinfo{pages}{83–124}.
\newblock \DOIprefix\doi{10.1017/jfm.2018.476}.
\bibitem[{Carlberg et~al.(2017)Carlberg, Barone and Antil}]{galerkin_problem}
\bibinfo{author}{Carlberg, K.}, \bibinfo{author}{Barone, M.}, \bibinfo{author}{Antil, H.}, \bibinfo{year}{2017}.
\newblock \bibinfo{title}{Galerkin v. least-squares petrov–galerkin projection in nonlinear model reduction}.
\newblock \bibinfo{journal}{Journal of Computational Physics} \bibinfo{volume}{330}, \bibinfo{pages}{693--734}.
\newblock \URLprefix \url{https://www.sciencedirect.com/science/article/pii/S0021999116305319}, \DOIprefix\doi{https://doi.org/10.1016/j.jcp.2016.10.033}.
\bibitem[{Corrochano et~al.(2023)Corrochano, Freitas, Parente and Clainche}]{corrochano_etal2023_NNComb}
\bibinfo{author}{Corrochano, A.}, \bibinfo{author}{Freitas, R.S.M.}, \bibinfo{author}{Parente, A.}, \bibinfo{author}{Clainche, S.L.}, \bibinfo{year}{2023}.
\newblock \bibinfo{title}{A predictive physics-aware hybrid reduced order model for reacting flows}.
\newblock \URLprefix \url{https://arxiv.org/abs/2301.09860}, \href{http://arxiv.org/abs/2301.09860}{{\tt arXiv:2301.09860}}.
\bibitem[{Ekambaram et~al.(2023)Ekambaram, Jati, Nguyen, Sinthong and Kalagnanam}]{temporal_mixer}
\bibinfo{author}{Ekambaram, V.}, \bibinfo{author}{Jati, A.}, \bibinfo{author}{Nguyen, N.}, \bibinfo{author}{Sinthong, P.}, \bibinfo{author}{Kalagnanam, J.}, \bibinfo{year}{2023}.
\newblock \bibinfo{title}{Tsmixer: Lightweight mlp-mixer model for multivariate time series forecasting}, in: \bibinfo{booktitle}{Proceedings of the 29th ACM SIGKDD Conference on Knowledge Discovery and Data Mining}, \bibinfo{publisher}{ACM}. p. \bibinfo{pages}{459–469}.
\newblock \URLprefix \url{http://dx.doi.org/10.1145/3580305.3599533}, \DOIprefix\doi{10.1145/3580305.3599533}.
\bibitem[{Gao et~al.(2024)Gao, Kaltenbach and Koumoutsakos}]{2024_gao_etal_gled}
\bibinfo{author}{Gao, H.}, \bibinfo{author}{Kaltenbach, S.}, \bibinfo{author}{Koumoutsakos, P.}, \bibinfo{year}{2024}.
\newblock \bibinfo{title}{Generative learning for forecasting the dynamics of high-dimensional complex systems}.
\newblock \bibinfo{journal}{Nature Communications} \bibinfo{volume}{15}, \bibinfo{pages}{8904}.
\newblock \URLprefix \url{https://doi.org/10.1038/s41467-024-53165-w}, \DOIprefix\doi{10.1038/s41467-024-53165-w}.
\bibitem[{Glorot and Bengio(2010)}]{xavier_initialization}
\bibinfo{author}{Glorot, X.}, \bibinfo{author}{Bengio, Y.}, \bibinfo{year}{2010}.
\newblock \bibinfo{title}{Understanding the difficulty of training deep feedforward neural networks}, in: \bibinfo{editor}{Teh, Y.W.}, \bibinfo{editor}{Titterington, M.} (Eds.), \bibinfo{booktitle}{Proceedings of the Thirteenth International Conference on Artificial Intelligence and Statistics}, \bibinfo{publisher}{PMLR}, \bibinfo{address}{Chia Laguna Resort, Sardinia, Italy}. pp. \bibinfo{pages}{249--256}.
\newblock \URLprefix \url{https://proceedings.mlr.press/v9/glorot10a.html}.
\bibitem[{Goodfellow et~al.(2016)Goodfellow, Bengio and Courville}]{2016_goodfellow_etal_book}
\bibinfo{author}{Goodfellow, I.}, \bibinfo{author}{Bengio, Y.}, \bibinfo{author}{Courville, A.}, \bibinfo{year}{2016}.
\newblock \bibinfo{title}{Deep Learning}.
\newblock \bibinfo{publisher}{MIT Press}.
\newblock \bibinfo{note}{\url{http://www.deeplearningbook.org}}.
\bibitem[{Han et~al.(2019)Han, Wang, Zhang and Chen}]{2019_han_etal_forecasting_convAutoEnc}
\bibinfo{author}{Han, R.}, \bibinfo{author}{Wang, Y.}, \bibinfo{author}{Zhang, Y.}, \bibinfo{author}{Chen, G.}, \bibinfo{year}{2019}.
\newblock \bibinfo{title}{{A novel spatial-temporal prediction method for unsteady wake flows based on hybrid deep neural network}}.
\newblock \bibinfo{journal}{Physics of Fluids} \bibinfo{volume}{31}, \bibinfo{pages}{127101}.
\newblock \URLprefix \url{https://doi.org/10.1063/1.5127247}, \DOIprefix\doi{10.1063/1.5127247}, \href{http://arxiv.org/abs/https://pubs.aip.org/aip/pof/article-pdf/doi/10.1063/1.5127247/13665230/127101\_1\_online.pdf}{{\tt arXiv:https://pubs.aip.org/aip/pof/article-pdf/doi/10.1063/1.5127247/13665230/127101\_1\_online.pdf}}.
\bibitem[{Hasegawa et~al.(2020)Hasegawa, Fukami, Murata and Fukagata}]{2020_hasegawa_etal_forecasting_convAutoEnc_2D}
\bibinfo{author}{Hasegawa, K.}, \bibinfo{author}{Fukami, K.}, \bibinfo{author}{Murata, T.}, \bibinfo{author}{Fukagata, K.}, \bibinfo{year}{2020}.
\newblock \bibinfo{title}{Machine-learning-based reduced-order modeling for unsteady flows around bluff bodies of various shapes}.
\newblock \bibinfo{journal}{Theoretical and Computational Fluid Dynamics} \bibinfo{volume}{34}, \bibinfo{pages}{367–383}.
\newblock \URLprefix \url{http://dx.doi.org/10.1007/s00162-020-00528-w}, \DOIprefix\doi{10.1007/s00162-020-00528-w}.
\bibitem[{He et~al.(2015)He, Zhang, Ren and Sun}]{kaiming_initialization}
\bibinfo{author}{He, K.}, \bibinfo{author}{Zhang, X.}, \bibinfo{author}{Ren, S.}, \bibinfo{author}{Sun, J.}, \bibinfo{year}{2015}.
\newblock \bibinfo{title}{Delving deep into rectifiers: Surpassing human-level performance on imagenet classification}, in: \bibinfo{booktitle}{2015 IEEE International Conference on Computer Vision (ICCV)}, pp. \bibinfo{pages}{1026--1034}.
\newblock \DOIprefix\doi{10.1109/ICCV.2015.123}.
\bibitem[{Hochreiter and Schmidhuber(1997)}]{hochreiter_etal_1997_lstm}
\bibinfo{author}{Hochreiter, S.}, \bibinfo{author}{Schmidhuber, J.}, \bibinfo{year}{1997}.
\newblock \bibinfo{title}{{Long Short-Term Memory}}.
\newblock \bibinfo{journal}{Neural Computation} \bibinfo{volume}{9}, \bibinfo{pages}{1735--1780}.
\newblock \URLprefix \url{https://doi.org/10.1162/neco.1997.9.8.1735}, \DOIprefix\doi{10.1162/neco.1997.9.8.1735}.
\bibitem[{Holmes et~al.(1996)Holmes, Lumley and Berkooz}]{holmes_lumley_libro}
\bibinfo{author}{Holmes, P.}, \bibinfo{author}{Lumley, J.L.}, \bibinfo{author}{Berkooz, G.}, \bibinfo{year}{1996}.
\newblock \bibinfo{title}{Turbulence, Coherent Structures, Dynamical Systems and Symmetry}.
\newblock Cambridge Monographs on Mechanics, \bibinfo{publisher}{Cambridge University Press}.
\bibitem[{Jiménez(2018)}]{Jimenez_2018}
\bibinfo{author}{Jiménez, J.}, \bibinfo{year}{2018}.
\newblock \bibinfo{title}{Coherent structures in wall-bounded turbulence}.
\newblock \bibinfo{journal}{Journal of Fluid Mechanics} \bibinfo{volume}{842}, \bibinfo{pages}{P1}.
\newblock \DOIprefix\doi{10.1017/jfm.2018.144}.
\bibitem[{Kingma and Ba(2017)}]{kingma2017adam}
\bibinfo{author}{Kingma, D.P.}, \bibinfo{author}{Ba, J.}, \bibinfo{year}{2017}.
\newblock \bibinfo{title}{Adam: A method for stochastic optimization}.
\newblock \URLprefix \url{https://arxiv.org/abs/1412.6980}, \href{http://arxiv.org/abs/1412.6980}{{\tt arXiv:1412.6980}}.
\bibitem[{Kičić et~al.(2023)Kičić, Vlachas, Arampatzis, Chatzimanolakis, Guibas and Koumoutsakos}]{adaled_petros}
\bibinfo{author}{Kičić, I.}, \bibinfo{author}{Vlachas, P.R.}, \bibinfo{author}{Arampatzis, G.}, \bibinfo{author}{Chatzimanolakis, M.}, \bibinfo{author}{Guibas, L.}, \bibinfo{author}{Koumoutsakos, P.}, \bibinfo{year}{2023}.
\newblock \bibinfo{title}{Adaptive learning of effective dynamics for online modeling of complex systems}.
\newblock \bibinfo{journal}{Computer Methods in Applied Mechanics and Engineering} \bibinfo{volume}{415}, \bibinfo{pages}{116204}.
\newblock \URLprefix \url{https://www.sciencedirect.com/science/article/pii/S0045782523003286}, \DOIprefix\doi{https://doi.org/10.1016/j.cma.2023.116204}.
\bibitem[{Kochkov et~al.(2024)Kochkov, Yuval, Langmore, Norgaard, Smith, Mooers, Kl{\"o}wer, Lottes, Rasp, D{\"u}ben, Hatfield, Battaglia, Sanchez-Gonzalez, Willson, Brenner and Hoyer}]{2024_weather_forecast}
\bibinfo{author}{Kochkov, D.}, \bibinfo{author}{Yuval, J.}, \bibinfo{author}{Langmore, I.}, \bibinfo{author}{Norgaard, P.}, \bibinfo{author}{Smith, J.}, \bibinfo{author}{Mooers, G.}, \bibinfo{author}{Kl{\"o}wer, M.}, \bibinfo{author}{Lottes, J.}, \bibinfo{author}{Rasp, S.}, \bibinfo{author}{D{\"u}ben, P.}, \bibinfo{author}{Hatfield, S.}, \bibinfo{author}{Battaglia, P.}, \bibinfo{author}{Sanchez-Gonzalez, A.}, \bibinfo{author}{Willson, M.}, \bibinfo{author}{Brenner, M.P.}, \bibinfo{author}{Hoyer, S.}, \bibinfo{year}{2024}.
\newblock \bibinfo{title}{Neural general circulation models for weather and climate}.
\newblock \bibinfo{journal}{Nature} \bibinfo{volume}{632}, \bibinfo{pages}{1060--1066}.
\newblock \URLprefix \url{https://doi.org/10.1038/s41586-024-07744-y}, \DOIprefix\doi{10.1038/s41586-024-07744-y}.
\bibitem[{Lakshminarayanan et~al.(2017)Lakshminarayanan, Pritzel and Blundell}]{uq_adaled}
\bibinfo{author}{Lakshminarayanan, B.}, \bibinfo{author}{Pritzel, A.}, \bibinfo{author}{Blundell, C.}, \bibinfo{year}{2017}.
\newblock \bibinfo{title}{Simple and scalable predictive uncertainty estimation using deep ensembles}.
\newblock \URLprefix \url{https://arxiv.org/abs/1612.01474}, \href{http://arxiv.org/abs/1612.01474}{{\tt arXiv:1612.01474}}.
\bibitem[{{Le Clainche} et~al.(2018){Le Clainche}, Pérez and Vega}]{LeClainche_2018_cilind}
\bibinfo{author}{{Le Clainche}, S.}, \bibinfo{author}{Pérez, J.M.}, \bibinfo{author}{Vega, J.M.}, \bibinfo{year}{2018}.
\newblock \bibinfo{title}{Spatio-temporal flow structures in the three-dimensional wake of a circular cylinder}.
\newblock \bibinfo{journal}{Fluid Dynamics Research} \bibinfo{volume}{50}, \bibinfo{pages}{051406}.
\newblock \URLprefix \url{https://dx.doi.org/10.1088/1873-7005/aab2f1}, \DOIprefix\doi{10.1088/1873-7005/aab2f1}.
\bibitem[{Le~Clainche et~al.(2017)Le~Clainche, Varas and Vega}]{2017_leClainche_etal_adaptive_oil}
\bibinfo{author}{Le~Clainche, S.}, \bibinfo{author}{Varas, F.}, \bibinfo{author}{Vega, J.M.}, \bibinfo{year}{2017}.
\newblock \bibinfo{title}{Accelerating oil reservoir simulations using pod on the fly}.
\newblock \bibinfo{journal}{International Journal for Numerical Methods in Engineering} \bibinfo{volume}{110}, \bibinfo{pages}{79--100}.
\newblock \URLprefix \url{https://onlinelibrary.wiley.com/doi/abs/10.1002/nme.5356}, \DOIprefix\doi{https://doi.org/10.1002/nme.5356}, \href{http://arxiv.org/abs/https://onlinelibrary.wiley.com/doi/pdf/10.1002/nme.5356}{{\tt arXiv:https://onlinelibrary.wiley.com/doi/pdf/10.1002/nme.5356}}.
\bibitem[{Le~Clainche and Vega(2017)}]{2017_leClainche_etal_hodmd}
\bibinfo{author}{Le~Clainche, S.}, \bibinfo{author}{Vega, J.M.}, \bibinfo{year}{2017}.
\newblock \bibinfo{title}{{Higher order dynamic mode decomposition to identify and extrapolate flow patterns}}.
\newblock \bibinfo{journal}{Physics of Fluids} \bibinfo{volume}{29}, \bibinfo{pages}{084102}.
\newblock \URLprefix \url{https://doi.org/10.1063/1.4997206}, \DOIprefix\doi{10.1063/1.4997206}, \href{http://arxiv.org/abs/https://pubs.aip.org/aip/pof/article-pdf/doi/10.1063/1.4997206/16002949/084102\_1\_online.pdf}{{\tt arXiv:https://pubs.aip.org/aip/pof/article-pdf/doi/10.1063/1.4997206/16002949/084102\_1\_online.pdf}}.
\bibitem[{Lin et~al.(2024)Lin, Li, Li, Li and Gao}]{Lin_etal_2024_diffusion_forecasting_survey}
\bibinfo{author}{Lin, L.}, \bibinfo{author}{Li, Z.}, \bibinfo{author}{Li, R.}, \bibinfo{author}{Li, X.}, \bibinfo{author}{Gao, J.}, \bibinfo{year}{2024}.
\newblock \bibinfo{title}{Diffusion models for time-series applications: a survey}.
\newblock \bibinfo{journal}{Frontiers of Information Technology {\&} Electronic Engineering} \bibinfo{volume}{25}, \bibinfo{pages}{19--41}.
\newblock \URLprefix \url{https://doi.org/10.1631/FITEE.2300310}, \DOIprefix\doi{10.1631/FITEE.2300310}.
\bibitem[{Loshchilov and Hutter(2017)}]{lr_schedule}
\bibinfo{author}{Loshchilov, I.}, \bibinfo{author}{Hutter, F.}, \bibinfo{year}{2017}.
\newblock \bibinfo{title}{Sgdr: Stochastic gradient descent with warm restarts}.
\newblock \URLprefix \url{https://arxiv.org/abs/1608.03983}, \href{http://arxiv.org/abs/1608.03983}{{\tt arXiv:1608.03983}}.
\bibitem[{Luchtenburg et~al.(2009)Luchtenburg, Noack and Schlegel}]{2009_luchtenburg_etal_galerkin}
\bibinfo{author}{Luchtenburg, D.}, \bibinfo{author}{Noack, B.}, \bibinfo{author}{Schlegel, M.}, \bibinfo{year}{2009}.
\newblock \bibinfo{title}{Galerkin method for fluid flows with analytical examples and matlab source codes.}
\newblock \bibinfo{journal}{Berlin Institute of Technology. Report 01.} .
\bibitem[{Lumley(1967)}]{lumley_pod}
\bibinfo{author}{Lumley, J.L.}, \bibinfo{year}{1967}.
\newblock \bibinfo{title}{The structure of inhomogeneous turbulent flows}, in: \bibinfo{editor}{Yaglom, A.M.}, \bibinfo{editor}{Tartarsky, V.I.} (Eds.), \bibinfo{booktitle}{Atmospheric Turbulence and Radio Wave Propagation}, \bibinfo{publisher}{Nauka}, \bibinfo{address}{Moscow}. pp. \bibinfo{pages}{166--178}.
\bibitem[{Makridakis et~al.(2018)Makridakis, Spiliotis and Assimakopoulos}]{MAKRIDAKIS_etal_2018_M4}
\bibinfo{author}{Makridakis, S.}, \bibinfo{author}{Spiliotis, E.}, \bibinfo{author}{Assimakopoulos, V.}, \bibinfo{year}{2018}.
\newblock \bibinfo{title}{The m4 competition: Results, findings, conclusion and way forward}.
\newblock \bibinfo{journal}{International Journal of Forecasting} \bibinfo{volume}{34}, \bibinfo{pages}{802--808}.
\newblock \URLprefix \url{https://www.sciencedirect.com/science/article/pii/S0169207018300785}, \DOIprefix\doi{https://doi.org/10.1016/j.ijforecast.2018.06.001}.
\bibitem[{Martínez-Sánchez et~al.(2023)Martínez-Sánchez, López, Le~Clainche, Lozano-Durán, Srivastava and Vinuesa}]{pod_40_energy_causality}
\bibinfo{author}{Martínez-Sánchez, A.}, \bibinfo{author}{López, E.}, \bibinfo{author}{Le~Clainche, S.}, \bibinfo{author}{Lozano-Durán, A.}, \bibinfo{author}{Srivastava, A.}, \bibinfo{author}{Vinuesa, R.}, \bibinfo{year}{2023}.
\newblock \bibinfo{title}{Causality analysis of large-scale structures in the flow around a wall-mounted square cylinder}.
\newblock \bibinfo{journal}{Journal of Fluid Mechanics} \bibinfo{volume}{967}, \bibinfo{pages}{A1}.
\newblock \DOIprefix\doi{10.1017/jfm.2023.423}.
\bibitem[{Mendez et~al.(2020)Mendez, Hess, Watz and Buchlin}]{Mendez_2020_experimental}
\bibinfo{author}{Mendez, M.A.}, \bibinfo{author}{Hess, D.}, \bibinfo{author}{Watz, B.B.}, \bibinfo{author}{Buchlin, J.M.}, \bibinfo{year}{2020}.
\newblock \bibinfo{title}{Multiscale proper orthogonal decomposition (mpod) of tr-piv data—a case study on stationary and transient cylinder wake flows}.
\newblock \bibinfo{journal}{Measurement Science and Technology} \bibinfo{volume}{31}, \bibinfo{pages}{094014}.
\newblock \URLprefix \url{https://dx.doi.org/10.1088/1361-6501/ab82be}, \DOIprefix\doi{10.1088/1361-6501/ab82be}.
\bibitem[{Mohan and Gaitonde(2018)}]{forecasting_pod_dl_2018}
\bibinfo{author}{Mohan, A.T.}, \bibinfo{author}{Gaitonde, D.V.}, \bibinfo{year}{2018}.
\newblock \bibinfo{title}{A deep learning based approach to reduced order modeling for turbulent flow control using lstm neural networks}.
\newblock \URLprefix \url{https://arxiv.org/abs/1804.09269}, \href{http://arxiv.org/abs/1804.09269}{{\tt arXiv:1804.09269}}.
\bibitem[{Nakamura et~al.(2021)Nakamura, Fukami, Hasegawa, Nabae and Fukagata}]{2021_nakamura_etal_forecasting_convAutoEnc_3D}
\bibinfo{author}{Nakamura, T.}, \bibinfo{author}{Fukami, K.}, \bibinfo{author}{Hasegawa, K.}, \bibinfo{author}{Nabae, Y.}, \bibinfo{author}{Fukagata, K.}, \bibinfo{year}{2021}.
\newblock \bibinfo{title}{{Convolutional neural network and long short-term memory based reduced order surrogate for minimal turbulent channel flow}}.
\newblock \bibinfo{journal}{Physics of Fluids} \bibinfo{volume}{33}, \bibinfo{pages}{025116}.
\newblock \URLprefix \url{https://doi.org/10.1063/5.0039845}, \DOIprefix\doi{10.1063/5.0039845}, \href{http://arxiv.org/abs/https://pubs.aip.org/aip/pof/article-pdf/doi/10.1063/5.0039845/13797759/025116\_1\_online.pdf}{{\tt arXiv:https://pubs.aip.org/aip/pof/article-pdf/doi/10.1063/5.0039845/13797759/025116\_1\_online.pdf}}.
\bibitem[{Noack et~al.(2011)Noack, Morzynski and Tadmor}]{2011_noack_etal_galerkin}
\bibinfo{author}{Noack, B.R.}, \bibinfo{author}{Morzynski, M.}, \bibinfo{author}{Tadmor, G.}, \bibinfo{year}{2011}.
\newblock \bibinfo{title}{Reduced-order modelling for flow control.}
\newblock \bibinfo{journal}{New York: Springer.} .
\bibitem[{Parish and Carlberg(2020a)}]{pod_forecating_2020_parish}
\bibinfo{author}{Parish, E.J.}, \bibinfo{author}{Carlberg, K.T.}, \bibinfo{year}{2020}a.
\newblock \bibinfo{title}{Time-series machine-learning error models for approximate solutions to parameterized dynamical systems}.
\newblock \bibinfo{journal}{Computer Methods in Applied Mechanics and Engineering} \bibinfo{volume}{365}, \bibinfo{pages}{112990}.
\newblock \URLprefix \url{https://www.sciencedirect.com/science/article/pii/S0045782520301742}, \DOIprefix\doi{https://doi.org/10.1016/j.cma.2020.112990}.
\bibitem[{Parish and Carlberg(2020b)}]{2020_parish_etal_LSTM_traditional}
\bibinfo{author}{Parish, E.J.}, \bibinfo{author}{Carlberg, K.T.}, \bibinfo{year}{2020}b.
\newblock \bibinfo{title}{Time-series machine-learning error models for approximate solutions to parameterized dynamical systems}.
\newblock \bibinfo{journal}{Computer Methods in Applied Mechanics and Engineering} \bibinfo{volume}{365}, \bibinfo{pages}{112990}.
\newblock \URLprefix \url{https://www.sciencedirect.com/science/article/pii/S0045782520301742}, \DOIprefix\doi{https://doi.org/10.1016/j.cma.2020.112990}.
\bibitem[{Pawar et~al.(2019)Pawar, Rahman, Vaddireddy, San, Rasheed and Vedula}]{forecasting_pod_dl_2019}
\bibinfo{author}{Pawar, S.}, \bibinfo{author}{Rahman, S.M.}, \bibinfo{author}{Vaddireddy, H.}, \bibinfo{author}{San, O.}, \bibinfo{author}{Rasheed, A.}, \bibinfo{author}{Vedula, P.}, \bibinfo{year}{2019}.
\newblock \bibinfo{title}{{A deep learning enabler for nonintrusive reduced order modeling of fluid flows}}.
\newblock \bibinfo{journal}{Physics of Fluids} \bibinfo{volume}{31}, \bibinfo{pages}{085101}.
\newblock \URLprefix \url{https://doi.org/10.1063/1.5113494}, \DOIprefix\doi{10.1063/1.5113494}, \href{http://arxiv.org/abs/https://pubs.aip.org/aip/pof/article-pdf/doi/10.1063/1.5113494/19764542/085101\_1\_online.pdf}{{\tt arXiv:https://pubs.aip.org/aip/pof/article-pdf/doi/10.1063/1.5113494/19764542/085101\_1\_online.pdf}}.
\bibitem[{Quarteroni et~al.(2016)Quarteroni, Manzoni and Negri}]{2016_quarteroni_etal_galerkin}
\bibinfo{author}{Quarteroni, A.}, \bibinfo{author}{Manzoni, A.}, \bibinfo{author}{Negri, F.}, \bibinfo{year}{2016}.
\newblock \bibinfo{title}{Reduced basis methods for partial differential equations.}
\newblock \bibinfo{journal}{Springer} .
\bibitem[{Rapún et~al.(2015)Rapún, Terragni and Vega}]{rapun_adaptive_pod_modes_2015}
\bibinfo{author}{Rapún, M.L.}, \bibinfo{author}{Terragni, F.}, \bibinfo{author}{Vega, J.M.}, \bibinfo{year}{2015}.
\newblock \bibinfo{title}{Adaptive pod-based low-dimensional modeling supported by residual estimates}.
\newblock \bibinfo{journal}{International Journal for Numerical Methods in Engineering} \bibinfo{volume}{104}, \bibinfo{pages}{844--868}.
\newblock \URLprefix \url{https://onlinelibrary.wiley.com/doi/abs/10.1002/nme.4947}, \DOIprefix\doi{https://doi.org/10.1002/nme.4947}, \href{http://arxiv.org/abs/https://onlinelibrary.wiley.com/doi/pdf/10.1002/nme.4947}{{\tt arXiv:https://onlinelibrary.wiley.com/doi/pdf/10.1002/nme.4947}}.
\bibitem[{Regazzoni et~al.(2019)Regazzoni, Dedè and Quarteroni}]{pod_forecating_2019_regazzoni}
\bibinfo{author}{Regazzoni, F.}, \bibinfo{author}{Dedè, L.}, \bibinfo{author}{Quarteroni, A.}, \bibinfo{year}{2019}.
\newblock \bibinfo{title}{Machine learning for fast and reliable solution of time-dependent differential equations}.
\newblock \bibinfo{journal}{Journal of Computational Physics} \bibinfo{volume}{397}, \bibinfo{pages}{108852}.
\newblock \URLprefix \url{https://www.sciencedirect.com/science/article/pii/S0021999119305364}, \DOIprefix\doi{https://doi.org/10.1016/j.jcp.2019.07.050}.
\bibitem[{Reiss et~al.(2018)Reiss, Schulze, Sesterhenn and Mehrmann}]{galerkin_low_num_modes}
\bibinfo{author}{Reiss, J.}, \bibinfo{author}{Schulze, P.}, \bibinfo{author}{Sesterhenn, J.}, \bibinfo{author}{Mehrmann, V.}, \bibinfo{year}{2018}.
\newblock \bibinfo{title}{The shifted proper orthogonal decomposition: A mode decomposition for multiple transport phenomena}.
\newblock \bibinfo{journal}{SIAM Journal on Scientific Computing} \bibinfo{volume}{40}, \bibinfo{pages}{A1322--A1344}.
\newblock \URLprefix \url{https://doi.org/10.1137/17M1140571}, \DOIprefix\doi{10.1137/17M1140571}, \href{http://arxiv.org/abs/https://doi.org/10.1137/17M1140571}{{\tt arXiv:https://doi.org/10.1137/17M1140571}}.
\bibitem[{Rempfer(2000)}]{Rempfer2000}
\bibinfo{author}{Rempfer, D.}, \bibinfo{year}{2000}.
\newblock \bibinfo{title}{On low-dimensional galerkin models for fluid flow}.
\newblock \bibinfo{journal}{Theoretical and Computational Fluid Dynamics} \bibinfo{volume}{14}, \bibinfo{pages}{75--88}.
\newblock \URLprefix \url{https://doi.org/10.1007/s001620050131}, \DOIprefix\doi{10.1007/s001620050131}.
\bibitem[{Sengupta et~al.(2025)Sengupta, Abadía-Heredia, Hetherington, Pérez and Clainche}]{2025_sengupta_eta_hybridML}
\bibinfo{author}{Sengupta, A.}, \bibinfo{author}{Abadía-Heredia, R.}, \bibinfo{author}{Hetherington, A.}, \bibinfo{author}{Pérez, J.M.}, \bibinfo{author}{Clainche, S.L.}, \bibinfo{year}{2025}.
\newblock \bibinfo{title}{Hybrid machine learning models based on physical patterns to accelerate cfd simulations: a short guide on autoregressive models}.
\newblock \URLprefix \url{https://arxiv.org/abs/2504.06774}, \href{http://arxiv.org/abs/2504.06774}{{\tt arXiv:2504.06774}}.
\bibitem[{Sirovich(1987)}]{sirovich_svd}
\bibinfo{author}{Sirovich, L.}, \bibinfo{year}{1987}.
\newblock \bibinfo{title}{Turbulence and the dynamics of coherent structures: I, ii, iii}.
\newblock \bibinfo{journal}{Quart. Appl. Math.} \bibinfo{volume}{45}, \bibinfo{pages}{561--590}.
\bibitem[{Stamatelopoulos and Sapsis(2025)}]{diffusion_models_extreme_events}
\bibinfo{author}{Stamatelopoulos, S.}, \bibinfo{author}{Sapsis, T.P.}, \bibinfo{year}{2025}.
\newblock \bibinfo{title}{Can diffusion models capture extreme event statistics?}
\newblock \bibinfo{journal}{Computer Methods in Applied Mechanics and Engineering} \bibinfo{volume}{435}, \bibinfo{pages}{117589}.
\newblock \URLprefix \url{https://www.sciencedirect.com/science/article/pii/S0045782524008430}, \DOIprefix\doi{https://doi.org/10.1016/j.cma.2024.117589}.
\bibitem[{Tsay(2001)}]{var_models_tsay_2001}
\bibinfo{author}{Tsay, R.}, \bibinfo{year}{2001}.
\newblock \bibinfo{title}{Analysis of Financial Time Series}.
\newblock \bibinfo{publisher}{John Wiley \& Sons}.
\bibitem[{Vega and {Le Clainche}(2021)}]{2021_leClainche_hodmd_book}
\bibinfo{author}{Vega, J.M.}, \bibinfo{author}{{Le Clainche}, S.}, \bibinfo{year}{2021}.
\newblock \bibinfo{title}{Higher Order Dynamic Mode Decomposition and Its Applications}.
\newblock \bibinfo{publisher}{Academic Press}.
\newblock \DOIprefix\doi{https://doi.org/10.1016/B978-0-12-819743-1.00007-0}.
\bibitem[{Vlachas et~al.(2018)Vlachas, Byeon, Wan, Sapsis and Koumoutsakos}]{2018_vlachas_etal_forecasting_chaos}
\bibinfo{author}{Vlachas, P.R.}, \bibinfo{author}{Byeon, W.}, \bibinfo{author}{Wan, Z.Y.}, \bibinfo{author}{Sapsis, T.P.}, \bibinfo{author}{Koumoutsakos, P.}, \bibinfo{year}{2018}.
\newblock \bibinfo{title}{Data-driven forecasting of high-dimensional chaotic systems with long short-term memory networks}.
\newblock \bibinfo{journal}{Proceedings of the Royal Society A: Mathematical, Physical and Engineering Sciences} \bibinfo{volume}{474}, \bibinfo{pages}{20170844}.
\newblock \URLprefix \url{https://royalsocietypublishing.org/doi/abs/10.1098/rspa.2017.0844}, \DOIprefix\doi{10.1098/rspa.2017.0844}, \href{http://arxiv.org/abs/https://royalsocietypublishing.org/doi/pdf/10.1098/rspa.2017.0844}{{\tt arXiv:https://royalsocietypublishing.org/doi/pdf/10.1098/rspa.2017.0844}}.
\bibitem[{Vreman(2004)}]{vreman_2004}
\bibinfo{author}{Vreman, A.W.}, \bibinfo{year}{2004}.
\newblock \bibinfo{title}{An eddy-viscosity subgrid-scale model for turbulent shear flow: Algebraic theory and applications}.
\newblock \bibinfo{journal}{Physics of Fluids} \bibinfo{volume}{16}, \bibinfo{pages}{3670--3681}.
\newblock \URLprefix \url{https://doi.org/10.1063/1.1785131}, \DOIprefix\doi{10.1063/1.1785131}, \href{http://arxiv.org/abs/https://pubs.aip.org/aip/pof/article-pdf/16/10/3670/19322749/3670\_1\_online.pdf}{{\tt arXiv:https://pubs.aip.org/aip/pof/article-pdf/16/10/3670/19322749/3670\_1\_online.pdf}}.
\bibitem[{Xiao et~al.(2019)Xiao, Heaney, Mottet, Fang, Lin, Navon, Guo, Matar, Robins and Pain}]{pod_dl_interpolation}
\bibinfo{author}{Xiao, D.}, \bibinfo{author}{Heaney, C.}, \bibinfo{author}{Mottet, L.}, \bibinfo{author}{Fang, F.}, \bibinfo{author}{Lin, W.}, \bibinfo{author}{Navon, I.}, \bibinfo{author}{Guo, Y.}, \bibinfo{author}{Matar, O.}, \bibinfo{author}{Robins, A.}, \bibinfo{author}{Pain, C.}, \bibinfo{year}{2019}.
\newblock \bibinfo{title}{A reduced order model for turbulent flows in the urban environment using machine learning}.
\newblock \bibinfo{journal}{Building and Environment} \bibinfo{volume}{148}, \bibinfo{pages}{323--337}.
\newblock \URLprefix \url{https://www.sciencedirect.com/science/article/pii/S0360132318306607}, \DOIprefix\doi{https://doi.org/10.1016/j.buildenv.2018.10.035}.
\bibitem[{Xu et~al.(2022)Xu, Zhou, Zhao, Guo and Zhang}]{forecasting_pod_dl_2022}
\bibinfo{author}{Xu, L.}, \bibinfo{author}{Zhou, G.}, \bibinfo{author}{Zhao, F.}, \bibinfo{author}{Guo, Z.}, \bibinfo{author}{Zhang, K.}, \bibinfo{year}{2022}.
\newblock \bibinfo{title}{A data-driven reduced order modeling for fluid flow analysis based on series forecasting intelligent algorithm}.
\newblock \bibinfo{journal}{IEEE Access} \bibinfo{volume}{10}, \bibinfo{pages}{60163--60176}.
\newblock \DOIprefix\doi{10.1109/ACCESS.2022.3177223}.

\end{thebibliography}

\end{document}